\documentclass{elsart}
\usepackage{epsfig}

\newcommand{\be}{\begin{eqnarray}}
\newcommand{\ee}{\end{eqnarray}}
\def\lsim{\mathrel{\rlap{\lower4pt\hbox{\hskip1pt$\sim$}}
	\raise1pt\hbox{$<$}}} %less than or approx. symbol
\def\gsim{\mathrel{\rlap{\lower4pt\hbox{\hskip1pt$\sim$}}
	\raise1pt\hbox{$>$}}} %greater than or approx. symbol
\def\kms{{km s$^{-1}$}}
\def\kpc{kpc}

\newcommand{\msun}{{M}_\odot}
\newcommand{\rsun}{{R}_\odot}

\begin{document}

\runauthor{Brown, Lee, Wijers, and Bethe}

\begin{frontmatter}
\title{Evolution of Black Holes in the Galaxy}
\author[suny]{G. E. Brown\thanksref{geb}}
\author[suny]{C.-H. Lee\thanksref{chl}}
\author[suny]{R.A.M.J. Wijers\thanksref{wijers}}
\author[cornell]{and H. A. Bethe}
\address[suny]{Department of Physics \& Astronomy,
        State University of New York,\\
        Stony Brook, New York 11794-3800, USA}
\address[cornell]{Floyd R. Newman Laboratory of Nuclear Studies,
        Cornell University, \\
        Ithaca, New York 14853, USA}
\thanks[geb]{E-mail: popenoe@nuclear.physics.sunysb.edu}
\thanks[chl]{E-mail: chlee@silver.physics.sunysb.edu}
\thanks[wijers]{E-mail: rwijers@astro.sunysb.edu}

\begin{abstract}
In this article we consider the formation and evolution of black holes,
especially those in binary stars where radiation from the matter falling
on them can be seen. We consider a number of effects introduced by some
of us, which are not traditionally included in binary evolution of massive 
stars. These are (i) hypercritical accretion, which allows neutron
stars to accrete enough matter to collapse to a black hole during their
spiral-in into another star. (ii) the strong mass loss of helium stars,
which causes their evolution to differ from that of the helium core
of a massive star. (iii) The direct formation of low-mass black holes
($M\sim2\msun$) from single stars, a consequence of a significant
strange-matter content of the nuclear-matter equation of state at high
density.  We discuss these processes here, and then review how they
affect various populations of binaries with black holes and neutron stars.

We have found that hypercritical accretion changes the standard scenario
for the evolution of binary neutron stars: it now usually gives a black-hole,
neutron-star (BH-NS) binary, because the first-born neutron star
collapses to a low-mass black hole in the course of the evolution.
A less probable double helium star scenario has to be introduced in
order to form neutron-star binaries.  The result is that low-mass
black-hole, neutron star (LBH-NS) binaries dominate the rate of detectable
gravity-wave events, say, by LIGO, by a factor $\sim 20$ over the binary
neutron stars.

The formation of high-mass black holes is suppressed somewhat due to the
influence of mass loss on the cores of massive stars, raising the
minimum mass for a star to form a massive BH to perhaps $80\,\msun$. Still,
inclusion of high-mass black-hole, neutron-star (HBH-NS) binaries increases
the predicted LIGO detection rate by another $\sim 30$ \%; lowering
of the mass loss rates of Wolf-Rayet stars may lower the HBH mass
limit, and thereby further increase the merger rate.

We predict that $\sim 33$ mergers per year will be observed with LIGO
once the advanced detectors planned to begin in 2004 are in place. 

Black holes are also considered as progenitors for gamma ray bursters
(GRB).  Due to their rapid spin, potentially high magnetic fields, and
relatively clean environment, mergers of black-hole, neutron-star binaries
may be especially suitable.  Combined with their ten times greater
formation rate than binary neutron stars this makes them attractive
candidates for GRB progenitors, although the strong concentration of
GRBs towards host galaxies may favor massive star progenitors or
helium-star, black-hole mergers.

We also consider binaries with a low-mass companion, and study the
evolution of the very large number of black-hole transients, consisting of
a black hole of mass $\sim 7\ \msun$ accompanied by a K or M main-sequence
star (except for two cases with a somewhat more massive subgiant donor).
We show that common envelope evolution must take place in the supergiant
stage of the massive progenitor of the black hole, giving an explanation
of why the donor masses are so small. We predict that there are about
22 times more binaries than observed, in which the main-sequence star,
somewhat more massive than a K- or M-star, sits quietly inside its Roche
Lobe, and will only become an X-ray source when the companion evolves
off the main sequence.

We briefly discuss the evolution of low-mass X-ray binaries into
millisecond pulsars. We point out that in the usual scenario for
forming millisecond pulsars with He white-dwarf companions, the long
period of stable mass transfer will usually lead to the collapse of
the neutron star into a black hole.  We then discuss Van den Heuvel's
``Hercules X-1 scenario" for forming low-mass X-ray binaries, commenting
on the differences in accretion onto the compact object by radiative or
semiconvective donors, rather than the deeply convective donors used in
the earlier part of our review.

In an appendix we describe the evolution of Cyg\,X-3, finding the
compact object to be a black hole of $\sim 3\ \msun$, together
with an $\sim 10\ \msun$ He star. In a second appendix we do the
accounting for gravitational mergers and in a third one we show
low-mass black-hole, neutron-star binaries to be good progenitors
for gamma ray bursters.

\end{abstract}

\end{frontmatter}

%-----------------------------------------------------------------------
\newpage
\tableofcontents

\newpage
\listoffigures
\listoftables

%-----------------------------------------------------------------------
\newpage
\section{Introduction}
\label{sec1}

The fate of massive stellar cores, both in single and binary stars,
has many observable consequences, both for what types of compact object
may be found in what type of binary, and for the formation rates of all
types of compact-object binary. We have discussed various aspects of
this problem in previous works, and here give an overview of all these
together, applying the same set of principles to all and obtaining a
consistent picture of the evolution of massive stars and binaries.

The best-known compact-object (i.e., neutron star or black hole) binaries
are the binary neutron stars. They are key testing grounds of general
relativity, and the usually favored gravity-wave source for LIGO.
Until recently the theoretical formation rate of binary neutron stars
gave at least one order-of-magnitude higher rate than was arrived at
empirically by extrapolation from observed binary neutron stars. Because
there are few binary neutron stars, and even fewer dominate the empirical
estimates, the latter are frequently revised. The recent doubling of
the estimated distance to PSR 1534$+$12 \cite{Stairs98} has lowered the
empirical birth rate significantly, widening the gap.

A solution to this discrepancy comes from combining the strange-matter
equation of state, which results in a relatively low maximum mass for
neutron stars, with hypercritical accretion \cite{BB98}.  In the standard
scenario the first neutron star formed spirals into the other star,
in a phase of common-envelope evolution. Bethe \& Brown \cite{BB98}
argued that when a neutron star spirals into a red giant with a deeply
convective envelope, it accretes matter at a very high rate of up to $1\
\msun$ yr$^{-1}$. Photons are trapped in the flow and carried
adiabatically inwards to the surface of the neutron star \cite{Brown95}.
The latter is heated to $T\sim 1$ MeV, temperatures at which neutrino
emission can carry off the thermal energy. Hence, the Eddington limit of
$\dot M_{\rm Edd}\sim 1.5\times 10^{-8}\ \msun$ yr$^{-1}$ does not apply.
As a result, the neutron star accretes about a solar mass of material and
collapses to a low-mass black hole. Only if the two stars are initially so
close in mass that at the time of spiral-in the first supernova has not
yet exploded (i.e.\ the object that spirals in is still a helium star)
a binary neutron star is formed. The sum total of binary neutron
stars and black-hole, neutron-star binaries is almost the same as what
was found for binary neutron stars in previous estimates, 
but now the binary neutron
stars are only a small fraction of the total.  The result is that an
order of magnitude more black-hole, neutron-star binaries than binary
neutron stars are formed.  Together with the fact that the black holes are
somewhat more massive than neutron stars, this implies that binaries with
black holes should play an important part in mergers producing gravitation
waves. They may also be good candidates for producing gamma-ray bursts.

No low-mass black-hole, neutron-star
binaries have been observed. This is due to the fact that the one
neutron star in them is unrecycled, hence is observable for only a 
short time. The rarer binary neutron stars, like PSR\,1913$+$16, do have
a long-lived recycled pulsar, which more than offsets their lower formation
rate and makes them dominate the observed population.

We do observe high-mass black holes in Cyg\,X-1 and in soft X-ray transients.
In the former, the black hole is of $\gsim 10\ \msun$ \cite{Herrero95}.
The companion O-star is near its Roche Lobe, and its wind is continuously
feeding the black hole, which shines through X-ray emission. In addition
to Cyg\,X-1, high-mass black holes are seen in the LMC in LMC\,X-3 and 
perhaps LMC\,X-1.
Much more copious are the transient sources, with black holes of mass
$M_{\rm BH}\sim 7\ \msun$, most of which flare up only occasionally with
long quiescent times between flare ups. Wijers \cite{Wijers96} estimated
$\sim 3000$ of these in the Galaxy. That is, these are the numbers that are 
presently operative.
Remarkable about the transient sources with unevolved donors is that the
main sequence star is K- or M-star, less massive than our sun.
Brown, Lee, \& Bethe \cite{BLB99} explain this in terms such that
higher-mass donors can also participate in the formation of such binaries
containing a high-mass black hole, but will end up in the evolution further 
away from the black hole so that they can pour matter on the latter only
when they evolve in subgiant or giant stage. Thus, there are a large factor
estimated to be $\sim 22$ more of those binaries which will not be seen
until the main sequence star evolves \cite{BLB99}.
The mechanism describing the evolution of the transient sources required
the massive progenitor of the black hole to carry out core helium burning
as if it were a single star; i.e., before having its H envelope removed
in RLOF by its main sequence companion. An interval of $\sim 20 -35\,\msun$
ZAMS was estimated for the progenitors of the high-mass black hole.
Consequently, this same interval of single stars, not in binary, would be
expected to end up as high-mass black holes. In the formation of these
high-mass black holes, most of the helium envelope of the progenitor
must drop into the black hole in order to form their high mass, so little
matter is returned to the Galaxy.

This brings us to the intriguing matter of SN 1987A which we believe did
go into a black hole, but after the supernova explosion which did return matter
to the Galaxy. The progenitor of SN 1987A was known to have ZAMS mass
$\sim 18\ \msun$. This leads us to the interesting concept of low-mass
black holes with delayed explosion, which result from the ZAMS mass range
$\sim 18-20\ \msun$, although the precise interval is uncertain. The
delayed explosion mechanism has been elucidated by Prakash et al. 
\cite{Prakash}. The range of ZAMS masses of single stars
in which neutron stars are formed is thus only $\sim 10-18\ \msun$.

The absence of matter being returned to the Galaxy in the ZAMS mass range
$\sim 20-35\ \msun$ impacts on nucleosynthesis, especially in the amount of
oxygen produced. Bethe \& Brown \cite{BB99} suggested that matter was again
returned to the Galaxy by stars in the ZAMS range $\sim 35-80\ \msun$.
In this case, the progenitor was stripped of H envelope in an LBV phase,
and the naked He star was suggested to evolve to a low-mass black hole,
with return of matter to the galaxy before its formation in a delayed
explosion, or to a neutron star. Thus, elements like oxygen were produced
in a bimodal distribution of ZAMS masses $M\lsim 20\ \msun$ and
$35\ \msun \lsim M \lsim 80\ \msun$.

The Bethe \& Brown \cite{BB99} suggestion was based on naked He stars
evolved by Woosley, Langer, \& Weaver \cite{WLW95} who used a too-large
wind loss rate for He stars. Wellstein \& Langer \cite{Wellstein99} have
evolved naked He stars with lower rates, in which case the final He
envelope is somewhat larger. However, the central carbon abundance
following core He burning is high $\sim 33$\%. With this abundance,
the stars will not skip the convective carbon burning stage in their
evolution, and according to the arguments of Brown, Lee, \& Bethe \cite{BLB99}
would still be expected to end up as low-mass compact objects, in which case
matter would be returned to the Galaxy. This matter will not, however, be
settled until the CO cores evolved with lowered He-star wind loss rates by
Wellstein \& Langer have been burned further up to the Fe core stage, so
the Bethe \& Brown \cite{BB99} bimodal mass region for nucleosynthesis
should be viewed as provisional. 

In sect.~\ref{sec2}, we discuss the maximum mass of neutron stars and
the processes that determine which range of initial stellar masses
gives rise to what compact object, and how mass loss in naked helium stars
changes those ranges. Then we describe the Bethe \& Brown \cite{BB98}
scenario for the evolution of massive binary stars, and especially
their treatment of common-envelope evolution and hypercritical accretion
(sect.~\ref{sec3}). We then discuss a few specific objects separately,
first binary neutron stars (sect.~\ref{sec4}), then Cyg\,X-1 and its ilk
(sect.~\ref{sec5}) and the black-hole transients (sect.~\ref{sec6}).
Then we comment briefly on how our results would affect the evolution
of low-mass X-ray binaries with neutron stars (sect.~\ref{sec7}) and
summarize our conclusions (sect.~\ref{sec8}). The discussion of Cyg\,X-3
and the possible implications of neutron-star, black-hole binaries
for gravity waves and gamma-ray bursts are in appendices 1--3.

%----------------------------------------------------------
\section{The Compact Star}
\label{sec2}

Thorsson, Prakash, \& Lattimer \cite{TPL94} and Brown \& Bethe \cite{BB94}
have studied the compact core after the collapse of a supernova, assuming
reasonable interactions between hadrons. Initially, the core consists
of neutrons, protons and electrons and a few neutrinos.
It has been called a proto-neutron star. It is stabilized against
gravity by the pressure of the Fermi gases of nucleons and leptons,
provided its mass is less than a limiting mass $M_{\rm PC}$
(proto-compact) of $\sim 1.8 \msun$.

If the assembled core mass is greater than $M_{\rm PC}$ there is no
stability and no bounce; the core collapses immediately into a black hole.
It is reasonable to take the core mass to be equal to the mass of the
Fe core in the pre-supernova, and we shall make this assumption,
although small corrections for fallback in the later supernova explosion
can be made as in Brown, Weingartner \& Wijers \cite{BWW96}. 
If the center collapses into a black hole, the outer part of the star 
has no support (other than centrifugal force from angular momentum) 
and will also collapse.

If the mass of the core is less than $M_{\rm PC}$, the electrons will be
captured by protons
    \be
    p+e^- \rightarrow n+\nu
    \ee
and the neutrinos will diffuse out of the core. This process takes of order
of 10 seconds, as has been shown by the duration of the neutrino signal
from SN 1987A. The result is a neutron star, with a small concentration of
protons and electrons.
The Fermi pressures of the core are chiefly from the nucleons, with small
correction from the electrons. On the other hand the nucleon energy is
increased by the symmetry energy;
i.e., by the fact that we now have nearly pure neutrons instead of
an approximately equal number of neutrons and protons.
Thorsson et al. \cite{TPL94} have calculated that the maximum mass of the
neutron star $M_{\rm NS}$ is still about $1.8\msun$; i.e., the symmetry
energy compensates the loss of the Fermi energy of the leptons.
Corrections for thermal pressure are small \cite{PBPELK97}.

The important fact is that the ten seconds of neutrino diffusion from
the core give ample time for the development of a shock which expels
most of the mass of the progenitor star.

But this is not the end of the story. The neutrons can convert into
protons plus $K^-$ mesons,
     \be
     n\rightarrow p+K^-.
     \label{eq2.2}
     \ee
This is short-hand for the more complicated interaction
$N+e^-\rightarrow N^\prime+K^-+\nu$ where $N$ is a nucleon. The neutrinos
leave the star. The times are sufficiently long that chemical
equilibrium is assured.
Since the density at the center of the neutron star is very high, the
energy of the $K^-$ is very low, as confirmed by Li, Lee, \& Brown \cite{LLB97}
using experimental data. By this conversion the nucleons can again become
half neutrons and half protons, thereby saving the symmetry energy needed
for pure neutron matter. The $K^-$, which are bosons, will condense,
saving the kinetic energy of the electrons they replace. The reaction
eq.~(\ref{eq2.2}) will be slow, since it is preceded by
     \be
     e^-\rightarrow K^- +\nu
     \label{eq2.3}
     \ee
(with the reaction eq.~(\ref{eq2.2}) following) as it becomes energetically
advantageous to replace the fermionic electrons by the bosonic $K^-$'s
at higher densities.
Initially the neutrino states in the neutron star are filled up to
the neutrino chemical potential with trapped neutrinos, and it takes some
seconds for them to leave the star.
These must leave before new neutrinos can be formed from the process
eq.~(\ref{eq2.3}). Thorsson et al. \cite{TPL94} have calculated that the maximum
mass of a star in which reaction eq.~(\ref{eq2.2}) has gone to completion is
    \be
    M_{\rm NP} \simeq 1.5\ \msun\ ,
    \ee
where the lower suffix NP denotes their nearly equal content of
neutrons  and protons, although we continue to use the usual name
``neutron star".
This is the maximum mass of neutron stars, which is to be compared with
the masses determined in binaries. The masses of 19 neutron star masses
determined in radio pulsars \cite{Thorsett99} are
consistent with this maximum mass.

The core mass $M_{\rm C}$ formed by the collapse of a supernova must 
therefore be compared to the two limiting masses, 
$M_{\rm PC}$ and $M_{\rm NP}$. If
    \be
    {\rm (I)} \;\; M_{\rm C} > M_{\rm PC}
    \ee
we get a high mass black hole.  If
    \be
    {\rm (II)} \;\; M_{\rm PC} > M_{\rm C} > M_{\rm NP}
    \ee
we get a low-mass black hole, of mass $M_{\rm C}$. Only if
    \be
    {\rm (III)} \;\; M_{\rm C} < M_{\rm NP}
    \ee
do we get a neutron (more precisely, ``nucleon") star from the SN.
Only in this case can we observe a pulsar. In cases (II) and (III)
we can see a supernova display. In case (I) we receive only initial neutrinos
from electrons captured in the collapse before $M_C$ becomes greater than
$M_{\rm PC}$ but no light would reach us. (Except perhaps if the new
black hole rotates rapidly enough to power an explosion, a mechanism 
proposed by MacFadyen and Woosley \cite{MacWoos98} for gamma-ray bursts.)

Woosley, Langer, \& Weaver \cite{WLW93} evolve massive stars with mass loss.
For stars in the ZAMS mass range $\sim 20-30\ \msun$, mass loss is relatively
unimportant and since $M_{\rm PC}\gsim 1.8\ \msun$ for this range, we find
from the earlier calculation of Woosley \& Weaver \cite{WW95} that most
of the single stars in this range will go into high-mass black holes.
Evolution of these stars in binaries is another matter. 
Timmes, Woosley, \& Weaver \cite{TWW96}, Brown, Weingartner, \& Wijers
\cite{BWW96}, and Wellstein \& Langer \cite{Wellstein99} find that
substantially smaller core masses result if the hydrogen envelope is taken
off in RLOF so that the helium star is naked when it burns. Thus, stars
of ZAMS masses $\sim 20-35\ \msun$ in such binaries evolve into low-mass
compact cores, black hole or neutron star. Woosley, Langer, \& Weaver 
\cite{WLW93} used helium-star wind loss rates which were too high by a factor 
$\sim 2-3$, but lower wind losses give only slightly larger He cores in the 
ZAMS mass range $\sim 20-35\ \msun$ \cite{Wellstein99} so our above conclusion
is unlikely to be reversed.

On the other hand, the fate of single stars in the ZAMS mass range 
$\sim 35-80\ \msun$ is uncertain. In the published Woosley,
Langer, \& Weaver \cite{WLW93} work with too high mass loss rate, so
much matter is blown away, first in LBV stage and later in W.-R. stage
that low-mass compact objects, black-hole or neutron-star, result
\cite{BWW96}. Bethe \& Brown \cite{BB99} attribute this to the fact
that convective carbon burning is not skipped in these stars.
In this stage a lot of entropy can be removed by $\nu\bar\nu$ emission,
so that a low-entropy, and therefore small, core results. In this
range, Wellstein \& Langer \cite{Wellstein99} find central $^{12}C$ abundances
of $33-35\%$ following He core burning, more than double the $\sim 15\%$
required for convective carbon core burning. Therefore, we believe that this
range of stars will still go into low-mass compact objects, even though
their final He cores are substantially larger because of the lower, more
correct, He-star wind mass loss rates used by Wellstein \& Langer 
\cite{Wellstein99}. However, this problem cannot be considered as settled
until the Wellstein \& Langer CO cores are burned up to the Fe core stage.
We will therefore not discuss the evolution of Cyg\,X-1 like objects,
high-mass black holes accompanied by sufficiently massive giant companion
so that they shine continuously in X-rays. It is not clear to us whether
LMC\,X-3, with a high-mass black hole and a B-star companion of roughly
equal mass,
has a history more like Cyg\,X-1 or like the transient black-hole 
binaries which we discuss below.

Bethe \& Brown \cite{BB99} took $80\,\msun$ as the lower mass limit for
high-mass black hole formation in binaries
which experience RLOF; i.e., in those for which helium core burning proceeds
in a naked helium star. Because of
our above discussion, we believe this mass limit may be too high, so that the
contributions from high-mass black-hole, neutron-star binaries were, if
anything, underestimated in their work. However, we will not know until the
CO cores obtained with better He-star mass loss rates are evolved further.

%----------------------------------------------------------
\section{Evolution of Binary Compact Objects}
\label{sec3}

\def\A{{\rm A}}
\def\B{{\rm B}}

We summarize the Bethe \& Brown \cite{BB98} evolution of binary compact
objects, paying special attention to their common envelope evolution.
In particular, we shall show that their schematic evolution should
be applicable to donors with deeply convective envelopes, whereas for
non-convective or semiconvective envelopes, such as encountered in the
evolution of low-mass X-ray binaries, their common envelope evolution
would not be expected to apply.

We call the star that is initially heavier star A, the other star B.  
We denote initial masses by subscript $i$, so we have masses
$M_{\A,i}$, $M_{\B,i}$.  We denote their ratio by $q$; thus
    \be
    q = M_{\B,i}/M_{\A,i} \leq 1~~.
    \label{eq3.1}
    \ee
Following Portegies Zwart \& Yungelson \cite{SimonYun98}, we assume that
$q$ is distributed uniformly between 0 and 1.  Likewise, we also assume
that ln $a$ is uniformly distributed, where $a$ is the semi-major axis
of their orbit.

However, we assume different limits for $a$ than Portegies Zwart \&
Yungelson \cite{SimonYun98}.
Initially both stars are massive main sequence stars, with radius at least
3 $\rsun$, so $a>6\ \rsun = 4 \times 10^6$ km.  At the other end of the scale,
we require $a< 4 \times 10^9$ km. We assume that 50\% of all stars are 
binaries with separations in this range (stars in wider binaries would
evolve as if they were single).
Then the fraction of binaries in a given interval of $\ln a$ is
   \be
   d \phi = d (\ln a)/7.
   \label{eq3.2}
   \ee

We assume that a star needs an initial mass of
   \be
   M>M_s=10\ \msun
   \label{eq3.3}
   \ee
in order to go supernova. Therefore, if $\alpha$ is the total rate of SNe,
the rate of SNe in mass interval $dM$ is given by
   \be
   d\alpha =\alpha n \left(\frac{M}{10\ \msun}\right)^{-n}
   \frac{dM}{M}
   \label{eq3.4}
   \ee
where we have used a power-law initial mass function with $n=1.5$ (close
to the Salpeter value $n=1.35$).
The birth rate of supernova systems was taken to be 
   \be
   \alpha=0.02\,{\rm yr}^{-1}
   \label{eq3.5}
   \ee
in the Galaxy. By a supernova system we mean a single star that goes supernova
(i.e., has $M_{\rm ZAMS}>10\,\msun$) or a close binary containing at least
one such star (close here means within the separation range
mentioned above).
Bethe \& Brown \cite{BB98} find that if the primary is massive enough to
go supernova, then there is an $\sim 50\%$ chance for the secondary to also
go supernova. This was calculated for a distributions flat in 
$q= M_{\B,i}/M_{\A,i}$. Therefore, the supernova rate in our notation would
be $1.25\alpha=0.025\,{\rm yr}^{-1}$.

Using the Cordes \& Chernoff \cite{CC97} distribution of kick velocities,
$43\%$ of the binaries were found to survive the first explosion. Thus,
at this stage, we are left with a birth rate of 
   \be
   R=0.02\times\frac{1}{2}\times\frac{1}{2}
        \times 0.43 \simeq 2\times 10^{-3}\ {\rm per\; yr}
   \label{eq3.6}
   \ee
for the formation of binaries consisting of a neutron star with a companion
massive enough to go supernova ($M>10\,\msun$).
The life time of such systems is the companion life time of  
$\sim 10^7$\,yr, but star A will be a pulsar for only $\sim 5\times 10^6$ \,yr
because it will spin down electromagnetically until it is no longer
observable.
{}From these numbers we estimate the number of such systems to be
$\sim 10^4$ in the Galaxy.

Since the pulsar is unrecycled, the expected number should be compared
with the detected population of active radio pulsars in the galaxy,
about $10^3$. This number should be multiplied by a factor of $1/2$ for
binarity, a further factor of $1/2$ for a binary in which both stars can
go supernova and the 0.43 for survival of the first explosion. This would
leave the large number $\sim 10^2$ if pulsars with massive companions
were as easily detected as single pulsars. In fact, only 2 are observed;
PSR\,1259$-$63 with a Be-star companion and PSR\,0045$-$73 with a 
B-star companion.
Stellar winds interfere with the radio pulses from these binaries,
obscuring the narrower ones. Doppler shifts also make these difficult to
observe. Nevertheless, the factor necessary to reduce their observability
is large. We return to the subject later.

At this stage we have an $\sim 1.4\ \msun$ neutron star with O or B-star
companion. We take the latter to have mass $\sim 15\ \msun$. The giant
has a He core containing some $30\%$ of its mass, surrounded by an 
envelope consisting mainly of H. We take the envelope to be deeply
convective,\footnote{ 
      The assumption that the envelope is deeply convective is essential
      for our later treatment of common envelope evolution with
      hypercritical accretion. Recent developments with non-convective
      or semiconvective donors show that the accretion rate is also
      highly super-Eddington, but still significantly less
      \cite{KingBegel99,BlandBegel99,Tauris99}. For very massive donors
      the rate is always highly super-Eddington.
      }
so the entropy is constant. The particles, nuclei and electrons, are
nonrelativistic and thus have $\gamma=5/3$. Therefore, the envelope forms
a polytrope of index $n=3/2$. Applegate \cite{Applegate98} shows that
the binding energy of the envelope is
   \be
   E\simeq 0.6 G M_\B^2 R^{-1}
   \label{eq3.7}
   \ee
where $R$ is the outer radius. In this formula the binding energy is
decreased $50\%$ by the kinetic energy, $E$ containing both effects. 

The major difference of the Bethe \& Brown calculations and of case
H of Portegies Zwart \& Yungelson \cite{SimonYun98} compared with
other work is the use of hypercritical accretion.
In a series of papers Chevalier \cite{Chevalier93,Chevalier96}
showed that once $\dot M$ exceeded
$\sim 10^4\ \dot M_{\rm Edd}$,
the photons were carried inwards in the adiabatic inflow,
onto the neutron star. The surface of the latter was heated sufficiently
that energy could be carried off by neutrino pairs. Brown \cite{Brown95}
reproduced Chevalier's results in analytical form. The idea has a 
much longer history:
Colgate \cite{Colgate71} showed already in 1971 that if neutrinos carry
off the bulk of the energy, accretion can proceed at a much greater rate
than Eddington. In 1972 Zeldovich et al. \cite{Zeldovich72}, before the
introduction of common envelope evolution, used hypercritical
accretion of a cloud onto a neutron star.
Bisnovatyi-Kogan \& Lamzin \cite{Bisno84} and Chevalier \cite{Chevalier93}
pointed out that during the common envelope phase of binary evolution,
photons would be trapped and accretion could occur at much higher rates,
and that neutron stars that go through this phase generally will go into
black holes.

We begin by considering the work done by the neutron star on the envelope
matter that it accretes. This will turn out to be only a fraction of the
total work, the rest coming from the production of the wake, but it 
illustrates simply our procedure. Taking the neutron star to be at rest,
the envelope matter is incident on it with the Keplerian velocity $v$.
The rate of accretion is given by the Bondi-Hoyle-Lyttleton theory
    \be
    \frac{d M_{\A}}{dt}=\pi\rho v R_{\rm ac}^2
   \label{eq3.8}
    \ee
where $\rho$ is the density of the B material, $v$ is its velocity
relative to the neutron star A, and $R_{\rm ac}$ is the accretion radius
   \be
   R_{\rm ac} =2 G M_{\A} v^{-2}.
   \label{eq3.9}
   \ee
The rate of change of momentum $P$ is
   \be
   \frac{dP}{dt} = v\frac{dM_{\A}}{dt},
   \label{eq3.10}
   \ee
the matter being brought to rest on the neutron star, and this is equal to
the force $F$. Consequently, the rate at which the  neutron star does
work in the material is
   \be
   \dot E = F v = v^2 \frac{d M_{\A}}{dt}.
   \label{eq3.11}
   \ee
Inclusion of the work done in creating the wake involves numerical 
calculations \cite{Shima85,Ruffert94,RuffArn94} with the result that
the coefficient of the right-hand side of eq.~(\ref{eq3.11}) is changed; i.e,
   \be
   \dot E = \left(\frac{c_d}{2}\right) v^2 \frac{d M_{\A}}{dt},
   \label{eq3.12}
   \ee
with $c_d\sim 6-8$ for our supersonic flow. It is, in fact, very important
that the wake plays such a large role, in that it's the fact that $c_d/2 >1$
(We consider $c_d/2$ to be $\gg 1$) that makes our 
later common envelope evolution
strongly nonconservative, the proportion of the total H-envelope mass
accreted onto the neutron star being relatively small.

In eq.~(\ref{eq3.12}) $v^2$ is the velocity of the B (giant) material
relative to A, the neutron star. This is given by
   \be
   v^2= G(M_{\A} +M_{\B})\ a^{-1}.
   \label{eq3.13}
   \ee

The interaction energy of A and B is
   \be
   E=\frac 12 G M_{\A} M_{\B} a^{-1}.
   \label{eq3.14}
   \ee
Since we know $M_{\B,i}$ and $M_{\B,f}$, the initial mass of B and
the mass of its He core, our unknown is $a_f$. We can obtain it by considering
   \be
   Y= M_{\B}  a^{-1}
   \label{eq3.15}
   \ee
as one variable, $M_{\A}$ as the other. Differentiating eq.~(\ref{eq3.14})
we have
   \be
   \dot E=\frac 12 G \left( {\dot M}_{\A} Y +M_{\A} \dot Y\right)
   \label{eq3.16}
   \ee
whereas combining eqs.~(\ref{eq3.12}) and (\ref{eq3.13}) and neglecting
$M_{\A}$ with respect to $M_{\B}$, we have
   \be
   \dot E = G \left(\frac{c_d}{2}\right) Y \ M_{\A}.
   \label{eq3.17}
   \ee
Thus, eqs.~(\ref{eq3.16}) and (\ref{eq3.17}) are equal, so we have
   \be
   \frac{\dot M_{\A}}{M_{\A}} = \frac{1}{(c_d-1)} \frac{\dot Y}{Y},
   \label{eq3.18}
   \ee
which can be integrated to give
   \be
   M_{\A} \propto  Y^{1/(c_d-1)} = Y^{1/5}
   \label{eq3.19}
   \ee
where we have chosen $c_d=6$ \cite{Shima85}. The final energy is then
   \be
   E_f=\frac 12 G M_{\A,i} Y_i \left(\frac{Y_f}{Y_i}\right)^{6/5}.
   \label{eq3.20}
   \ee

The binding energy $E_f$ of star A to star B serves to expel the
envelope of star B, whose initial binding energy is given by eq.~(\ref{eq3.7}).
Mass transfer begins at the Roche Lobe which lies at $\sim 0.6 \ a_i$
for the masses involved. However, star B expands rapidly in red giant stage
before the mass transfer can be completed. To keep the numbers easy
to compare with Bethe \& Brown \cite{BB98}, we use their approximation
of starting spiral-in when the giant's radius equals the orbital separation
rather than the Roche-lobe radius. Since for the large mass ratios 
considered here, $R_{\rm L}/a\sim0.5$ for the giant, this implies
we require $E_f$ of eq.~(\ref{eq3.20}) to be about twice the binding
energy (eq.~\ref{eq3.7}), i.e.
   \be
   E_f = \frac{0.6}{\alpha} G \frac{M_{\B,i}^2}{a_i}
    = 1.2 \ G \frac{M_{\B,i}^2}{a_i}.
   \label{eq3.21}
   \ee
(We set the common-envelope efficiency, $\alpha$,
to 0.5.) The ejected material of B is, therefore,
released with roughly the thermal energy it had in the envelope; in 
other words,
the thermal energy content of the star is not used to help expel it.
Inserting eq.~(\ref{eq3.21}) into eq.~(\ref{eq3.20}) yields
   \be
   \left(\frac{Y_f}{Y_i}\right)^{1.2} =2.4 \frac{M_{\B,i}}{M_{\A,i}}.
   \label{eq3.22}
   \ee
Star A is initially a neutron star, $M_{\A,i}=1.4\ \msun$. For star B
we assume $M_{\B,i}=15\ \msun$. Then eq.~(\ref{eq3.22}) yields
   \be
   \frac{Y_f}{Y_i}=15.
   \label{eq3.23}
   \ee
We use this to find the result of accretion, with the help of 
eq.~(\ref{eq3.19}),
   \be
   \frac{M_{\A,f}}{M_{\A,i}}=1.73
   \label{eq3.24}
   \ee
or
   \be
   M_{\A, f}=2.4\ \msun.
   \label{eq3.25}
   \ee
This is well above any of the modern limits for neutron star masses, so we
find that the neutron star has gone into a black hole.

Our conclusion is, then, that in the standard scenario for evolving
binary neutron stars, if the giant is deeply convective, accretion in
the common envelope phase will convert the neutron star into a black hole.

Star B, by losing its envelope, becomes a He star, We estimate that
   \be
   \frac{M_{\B,f}}{M_{\B,i}}\simeq 0.3.
   \label{eq3.26}
   \ee
The size of the orbit is determined by eq.~(\ref{eq3.15}),
   \be
   \frac{a_i}{a_f}=\frac{M_{\B,i}}{M_{\B,f}}\ \frac{Y_f}{Y_i} = 50.
   \label{eq3.27}
   \ee
The final distance between the stars $a_f$ should not be less than about
$10^{11}$ cm, so that the He star (mass $M_{\B,f}$) fits within its Roche
lobe next to the black hole of mass $M_{\A,f}$. 
Bethe \& Brown \cite{BB98} showed that if the black hole and the
neutron star resulting from the explosion of star B are to merge in a
Hubble time, then $a_f <3.8 \times 10^{11}$ (for circular orbits; correction
for eccentricity will be given later).
Therefore the initial distance of the two stars, after the first mass
exchange and the first supernova should be
   \be
   0.5 \times 10^{13} \ {\rm cm} < a_i <
   1.9 \times 10^{13} \ {\rm cm}
   \label{eq3.28}
   \ee

If the initial distribution of distances is $da/7a$, the probability of
finding $a$ between the limits of eq.~(\ref{eq3.28}) is
   \be
   P=18\%.
   \label{eq3.29}
   \ee
As noted earlier, 43\% of the binaries survive the first explosion, so the
combined probability is now
   \be
   P=8\%
   \label{eq3.30}
   \ee
for the survivors falling in the logarithmic interval in which they survive
coalescence, but are narrow enough to merge in a Hubble time. Our final
result, following from a birth rate of $10^{-2}$ binaries per year in which
one star goes supernova, half of which have both stars going supernova, is
   \be
   R=10^{-2} \times 0.5\times 0.08 \times 0.5 =2\times 10^{-4}\ {\rm yr}^{-1}
   \label{eq3.31}
   \ee
in the Galaxy. The final factor of 0.5 is the survival rate of the He-star,
neutron star binary, calculated by Monte Carlo methods.
Bethe \& Brown \cite{BB98} quoted $10^{-4}$ yr$^{-1}$, or half of this rate,
in order to take into account some effects not considered by them in which
the binary disappeared (e.g., Portegies Zwart and Verbunt \cite{PZVer96}). 

Our final rate is, then,
   \be
   R=10^{-4} {\rm yr}^{-1}{\rm\,galaxy}^{-1}.
   \label{eq3.32}
   \ee
Using our supernova rate of $0.025$ per year, which includes the case where 
both stars in the binary go supernova, we can convert this birth rate
to $0.004$ per supernova for comparison with other work.
Portegies Zwart \& Yungelson \cite{SimonYun98} in their case H,
which included hypercritical accretion, got 0.0036 per supernova, within
10\% of our value. Thus, the chief difference
between our result in eq.~(\ref{eq3.32}) and the $R=5.3\times 10^{-5}$ of
these authors is due to the different assumed SN rate.

In our above estimates we have assumed the second neutron star to be formed
to have a circular orbit of the same $a$ as its He-star progenitor. However,
eccentricity in its orbit leads to a value of $a_f$ substantially larger than
the $3.8\times 10^{11}$ cm used above as the maximum separation for merger.
In general, most of the final
binaries will have $e > 0.5$, with a heavy peak in the distribution
close to $e = 1$. The rise
occurs because preservation of the binary in the explosion is substantially
greater if the kick velocity is opposite to the orbital velocity before
explosion.  In this case the eccentricity $\epsilon$ is large.  The most
favorable situation is when the orbital and kick velocities are equal in
magnitude.  (See the figures in Wettig and Brown \cite{Wettig96}.)  
Eggleton \cite{Eggleton98} has
kindly furnished us with a useful interpolation formula for the increase.  The
factor by which to multiply the time for merger in circular orbits, is
   \be
   Z(e) \approx (1 - e^2)^{3.689 - 0.243e - 0.058e^2} ~~.
   \ee
This formula is accurate to about 1\% for $e \leq 0.99$.  Thus, if the
initial eccentricity is 0.7, the time to shrink the orbit to zero is about 10\%
of the time required if the initial eccentricity were zero for the same
initial period.  The maximum $a_f = 3.8 \times 10^{11}$ cm for circular orbits
would be increased by the fourth root of the decrease in time; i.e., up to $6.8
\times 10^{11}$ cm for this eccentricity.  The maximum $a_i$ in 
eq.~(\ref{eq3.28})
would go up to $3.4 \times 10^{13}$ cm, increasing the favorable logarithmic
interval by $\sim$ 40\%.
We have not introduced this correction because it is of the same general
size as the uncertainty in the supernova rate. However,
this correction gives us some comfort that our final numbers are not 
unreasonably large.

If we produce an order of magnitude more low-mass black hole, neutron
star binaries than binary neutron stars, the obvious question is why we
have not seen any. The neutron star in this object is ``fresh" (unrecycled)
so it would spin down into the graveyard of neutron stars in $\sim
5\times 10^6$ yr. The two relativistic binary pulsars we do see 1913$+$16
and 1534$+$12 have been recycled, have magnetic fields $B\sim 10^{10}$ G, 
two orders of magnitude less than a fresh pulsar, and will therefore
be seen for about 100 times longer than an unrecycled neutron star.
So even with a ten times higher birth rate, we should see ten times
fewer LBH-NS binaries than NS-NS binaries.
Furthermore, the binary with black hole will have a somewhat higher mass,
therefore greater Doppler shift, and therefore be harder to detect.
In view of the above, it is reasonable
that our low-mass black-hole, neutron-star binaries have not been observed,
but they should be actively looked for.

We should also calculate the rate of coalescences of the black hole with the
He star. These have been suggested by Fryer \& Woosley \cite{FryWoos98} as
candidate progenitors for the long time gamma ray bursters. Note that they
will occur for a range of $0.04 \times 10^{13} {\rm \ cm} 
<a_i < 0.5\times 10^{13}$ cm, a logarithmic interval double that of
eq.~(\ref{eq3.28}). Thus, the black-hole, He-star coalescence has a
probability
   \be
   P=36\%.
   \ee
Furthermore, this situation does not have the 50\% disruption in the final
explosion, so the black-hole, He-star coalescences occur with a total rate
of 4 times that of the black-hole, neutron-star mergers.

There has been much discussion in the literature of the difficulties in
common envelope evolution. We believe our model of deeply convective
giants and hypercritical accretion offers an ideal case. Of course, the
initiation of the common envelope evolution requires some attention, but
it can be modeled in a realistic way \cite{Rasio96}. As the giant
evolves across its Roche lobe, the compact object creates a tidal
bulge in the giant envelope, which follows the compact object, torquing it in.
As the convective giant loses mass, the envelope expands in order to keep
entropy constant. In Bondi-Hoyle-Lyttleton accretion, a density
$\rho_\infty\sim 10^{-13} \ {\rm g\ cm}^{-3}$ is sufficient with wind
velocities $\sim 1000$ km s$^{-1}$ in order to give accretion at the
Eddington rate. Thus to achieve $\dot M\sim 10^8 M_{\rm Edd}
\sim 1 \ \msun {\rm yr}^{-1}$ we need $\rho\sim 10^{-5}$ g cm$^{-3}$
which is found at $0.9\ R$, where $R$ is the radius of
the giant. At this rate of accretion, angular momentum, etc, are hardly
able to impede it appreciably. The total mass accreted onto the compact
object is $\sim 1\ \msun$, so the common envelope evolution has
dynamical time of years. As noted earlier, it is non-conservative.

%----------------------------------------------------------
\section{Evolution of Binary Neutron Stars}
\label{sec4}

Since the standard scenario of evolution of binary compact objects
ends up with low-mass black-hole, neutron-star binaries, another way
must be found to evolve neutron star binaries. In the double He-star
scenario was suggested by Brown \cite{Brown95} and developed
further by Wettig \& Brown \cite{Wettig96} the neutron star
avoids going through common envelope with a companion star. 
In this way the neutron star can avoid being converted into a black hole
by accretion. For two giants to burn He at the same time, they must be
within $\sim 5\%$ of each other in mass, the helium burning time
being $\sim 10\%$ of the main sequence life time, and stellar evolution time
going roughly with the inverse square of the mass. With a flat mass ratio
distribution, this happens in 5\% of all cases, making the ratio of
NS-NS to NS-LBH binaries 1:20. However, when the primary becomes an LBH,
only half the secondaries will be massive enough to form a NS, whereas
for the very close mass values of the double-He scenario this factor 2
loss does not occur. Thus, binary
neutron stars should be formed $10\%$ as often as low-mass black-hole, 
neutron-star binaries. This $10\%$ is nearly model independent because
everything else roughly scales.

The scenario goes as in Wettig \& Brown \cite{Wettig96}. The primary
O-star evolves transferring its H-envelope to the companion. Often,
this would lead to `rejuvenation' of the secondary, i.e.\ its evolution
would restart also from the ZAMS with the now higher total mass, and it would
make a much heavier core. However, here the core of the secondary has
evolved almost as far as the primary's core, so the core molecular weight
is much higher then that of the envelope. This prevents convection
in the core from extending into the new envelope to make the bigger
core, so no rejuvenation takes place\cite{Braun95}. 
Since $q \sim 1$, the first mass transfer
is nearly conservative. The second is not, so the 
two He-cores then share a common
H envelope, which they expel, while dropping to a lower final separation
$a_f$.

Following the explosion of the first He star, the companion He-star pours
wind matter onto the pulsar, bringing the magnetic field down and spinning
it up \cite{Brown95,Wettig96}.  The end result is two neutron stars of
very nearly equal mass, although wind accretion can change the mass two
or three percent.

The above scenario ends for He-star masses greater than 4 or $5\ \msun$,
corresponding to ZAMS masses greater than $\sim 16$ or $18\ \msun$.
However, less massive He stars evolve in the He shell-burning stage, and
a further mass transfer (Case C) can take place. The transfer of He to
the pulsar can again bring about a black hole, which Brown \cite{Brown97}
very roughly estimates to occur in $\sim 50\%$ of the double neutron star
binaries. This is roughly consistent with results of Fryer \& Kalogera
\cite{Kalog97}. Taking a rate of $R=10^{-4}$ per year per galaxy for the
low-mass black-hole, neutron-star binaries, we thus arrive at a birth rate
of
   \be
   R\simeq 5\times 10^{-6} \ {\rm per\; year\; per\; galaxy}
   \label{eq4.1}
   \ee
for binary neutron-star formation. However, the black holes formed in the
He shell burning evolution will not have accreted much mass and will have
about the same chirp mass as binary neutron stars
(see below) for gravitational merging.

Our best guess values eqs.~(\ref{eq3.32}) and (\ref{eq4.1})
thus give an $\sim 20$ to 1 ratio for formation of low-mass black-hole,
neutron-star binaries to binary neutron stars. The former are better
progenitors for gravitational waves from mergers because of their
higher masses and they have many advantages as progenitors of
gamma ray bursters \cite{BWLLB99}. Note that our estimated rate of
$R=5\times 10^{-6}$ per galaxy per year for binary neutron star formation
is consistent with the empirical rates discussed in our introduction.

%----------------------------------------------------------
\section{High-Mass Black-Hole O/B-star Binaries}
\label{sec5}

We will be brief in our review of these, because we believe the
evolution of these objects such as Cyg\,X-1, LMC\,X-1 and LMC\,X-3
to be less well understood than the low-mass black-hole, neutron-star
binaries.
Evolutionary calculations now proceeding by Alexander Heger, using
the CO cores evolved by Wellstein \& Langer \cite{Wellstein99}
should clarify this situation substantially.

Bethe \& Brown \cite{BB99} arrived at a limit of ZAMS mass $80\ \msun$
for stars in binaries to go into high-mass black holes (unless Case C
mass transfer takes place as we discuss in our next section). 
This limiting mass is much higher than other workers have used. It was
based on calculations of Woosley, Langer, \& Weaver \cite{WLW93} and was
so high because of very high mass loss rates used by these authors. With
more correct lower rates the limiting mass may come down, so the
Bethe \& Brown evolution should be viewed as giving a lower limits to the 
number of high-mass black-hole, O/B-star binaries. Their estimated
birth rate of about $3\times10^{-5}$ per galaxy per year does agree
reasonably well with the fact that only one such system is known in
the Galaxy. However, since even with a twice larger separation the accretion
rate of the black hole from the fast wind of the O star becomes small,
it is possible that substantially more systems with somewhat wider
orbits exist undetected, and that Cyg\,X-1 is the only one presently in
the (very short) phase of incipient Roche lobe overflow when it is bright.
Bethe \& Brown \cite{BB99} found this narrowness of the Cyg\,X-1 orbit
($17~\rsun$ according to Herrero et al. \cite{Herrero95}) to be
puzzling: the massive stars in the progenitor binary initially had to
fit within
their Roche Lobes, therefore a separation of at least double the current
$17~\rsun$ was needed. And most evolutionary effects from then on, such
as wind mass loss or supernova-like mass loss, would tend to widen the
orbit. Of
course, the orbit could be narrowed in Case A mass transfer (i.e.\
during the main sequence) since the
progenitor of the black hole was more massive than the present donor,
but it could not become so narrow that the present donor filled its Roche
lobe, and would widen again once the mass ratio became reversed and
widen further due to wind loss after the whole primary envelope was lost.

In any case, a binary as narrow as Cyg\,X-1 would coalesce in the common
envelope evolution once the O-star
companion of the massive black-hole goes into red giant phase, according to the
Bethe \& Brown \cite{BB99} estimates. Since the black hole in Cyg\,X-1 has
mass $\gsim 10~\msun$ and is probably the most massive black hole in
a binary observed in the Galaxy, in the Fryer \& Woosley \cite{FryWoos98}
model where the black hole ``eats" the W.-R. companion, such a coalescence
should produce the most energetic long-lasting gamma ray burster.
We are unable to evaluate the probability of Cyg\,X-1 like objects 
merging following common envelope evolution because we have been unable
to understand why Cyg\,X-1, before common envelope evolution, is so narrow.
The LBV, RSG, and WNL stages of W.-R. development are not
quantitatively understood. 

%Bethe \& Brown \cite{BB99} found that the closer of the Cyg\,X-1-like
%objects would, after the other star had exploded and left a neutron star,
%lead eventually to a merger of the black hole and the neutron star.
After the main sequence star in a Cyg X-1-like object explodes
and becomes a neutron star, according to Bethe \& Brown \cite{BB99}
the binary will eventually merge.
They estimated the contribution to the merger rate of these systems
to be $(4-6)\times 10^{-6}\ {\rm yr^{-1}\ galaxy^{-1}}$, however with
considerable uncertainty due to the fact that the evolution of Cyg\,X-1
itself is uncertain.  Lowering the mass limit for black-hole formation by
having lower mass loss rates would increase this number (e.g.\ a limit
of 40\,$\msun$ would increase the merger rate by a factor 5).

%----------------------------------------------------------
\section{The Formation of High-Mass Black Holes in Low-Mass
 X-ray Binaries}
\label{sec6}

\subsection{General}

Crucial to our discussion here is the fact that single stars evolve very
differently from stars in binaries that lose their H-envelope
\cite{BLB99,Wellstein99,BWW96,TWW96} either on
the main sequence (Case A) or in the giant phase (Case B). However, stars that
transfer mass or lose mass after core He burning (Case C) evolve, for our 
purposes, as single stars, because the He core is then exposed too close to
its death for wind mass loss to significantly alter its fate. 
Single stars above a ZAMS mass of about $20~\msun$ skip convective
carbon burning following core He burning, with the result, as we shall
explain, that their Fe cores
are substantially more massive than stars in binaries, in which 
H-envelope has been transferred or lifted off before He core burning.
These latter ``naked" He stars burn $^{12}C$ convectively, 
and end up with relatively
small Fe cores. The reason that they do this has to do chiefly with the
large mass loss rates of the ``naked" He cores, which behave like W.-R.'s.
Unfortunately, in calculation until recently, substantially too large
mass loss rates were used, so we cannot pin limits down quantitatively.
In this section we will deal with the ZAMS mass range $\sim 20-35~\msun$,
in which it is clear that many, if not most, of the single stars go into 
high-mass black holes, whereas stars in binaries which burn ``naked" He
cores go into low-mass compact objects. In this region of
ZAMS masses the use of too-high He-star mass loss rates does not cause
large effects \cite{BLB99}.

The convective carbon burning phase (when it occurs) is
extremely important in pre-supernova evolution,
because this is the first phase in which a large amount of entropy
can be carried off in $\nu\bar\nu$-pair emission, especially if
this phase is of long duration.  The reaction in which carbon burns is
$^{12}C(\alpha,\gamma)^{16}O$ (other reactions like $C+C$ would require
excessive temperatures).  The cross section of $^{12}C(\alpha,\gamma)^{16}O$ is
still not accurately determined; the lower this cross section the higher the
temperature of the $^{12}C$ burning, and therefore the more intense the
$\nu\bar\nu$ emission.
With the relatively low
$^{12}C(\alpha,\gamma)^{16}O$ rates determined both directly from nuclear
reactions and from nucleosynthesis by Weaver \& Woosley \cite{WeaW93}, the
entropy carried off during $^{12}C$ burning in the stars of ZAMS
mass $\sim 10- 20~\msun$ is substantial.
The result is rather low-mass Fe cores for these stars, which can evolve
into neutron stars. Note that in the literature earlier than
Weaver \& Woosley \cite{WeaW93} often large $^{12}C(\alpha,\gamma)^{16}O$
rates were used, so that the $^{12}C$ was converted into oxygen and the
convective burning did not have time to be effective.
Thus its role was not widely appreciated.

Of particular importance is the ZAMS mass at which the convective carbon
burning is skipped.
%----------------
In the Woosley \& Weaver \cite{WW95} calculations this occurs at
ZAMS mass $19\msun$ but with a slightly lower $^{12}C(\alpha,\gamma)^{16}O$
rate it might come at $20\msun$ or higher \cite{Brown97}. As the progenitor
mass increases, it follows from general polytropic arguments that the
entropy at a given burning stage increases.
At the higher entropies of the more massive stars the density at which
burning occurs is lower, because the temperature is almost fixed for a
given fuel. Lower densities decrease the rate of the triple-$\alpha$
process which produces $^{12}C$ relative to the two-body
$^{12}C(\alpha,\gamma)^{16}O$ which produces oxygen.
Therefore, at the higher entropies in the more massive stars the
ratio of $^{12}C$ to $^{16}O$ at the end of He burning is lower.
The star skips the long convective carbon
burning and goes on to the much shorter oxygen burning.  Oxygen burning goes
via $^{16}O + ^{16}O$ giving various products, at very much higher temperature
than $C(\alpha,\gamma)$ and much faster.
Since neutrino cooling during the long carbon-burning phase gets
rid of a lot of entropy of the core, skipping this phase leaves
the core entropy higher and the final Chandrasekhar core fatter.

\begin{figure} %[ht]
\centerline{\epsfig{file=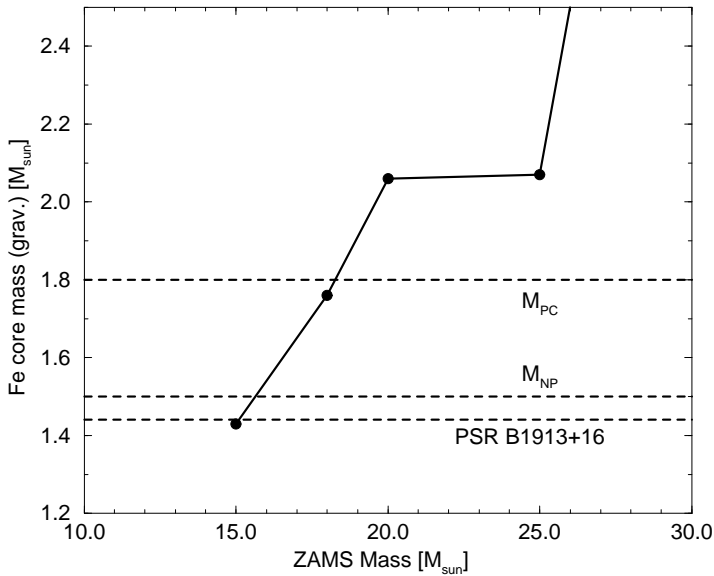,height=9cm}}
\caption{Compact core masses resulting from the
evolution of single stars, Case B of solar metallicity of Woosley \& Weaver 
\cite{WW95}.
The horizontal dashed lines indicate the mass of the heaviest known 
well--measured pulsar \cite{Thorsett99}, 
the maximum mass of a neutron star, and our
estimate of $M_{\rm PC}$ (proto-compact), the maximum compact core mass
for which matter can be returned to the galaxy.}
\label{fig1}
\end{figure}

In Fig.~\ref{fig1} the large jump in compact object mass in single stars at
ZAMS mass $\sim 19~\msun$ is clearly seen. From our discussion in
Section \ref{sec2} we see that this is just at the point where
the Fe core mass goes above the proto-compact mass of $\sim 1.8\msun$ and,
therefore, above this mass one would expect single stars to go into
high-mass black holes. Arguments have been given that SN 1987A
with progenitor ZAMS mass of $\sim 18~\msun$ evolved into a low-mass
black hole \cite{BB94}. We believe from our above arguments
and Fig.~\ref{fig1} that just above the ZAMS mass of $\sim 20~\msun$, single
stars go into high-mass black holes without return of matter to the
Galaxy.
Thus, the region of masses for low-mass black hole formation in
single stars is narrow, say $\sim 18 -20~\msun$ (although we believe
it to be much larger in binaries).

Thus far our discussion has been chiefly about single stars, in which
the He burns ``clothed" by a hydrogen envelope. In this case the
convective helium core grows in stars as time passes.
In the ``naked" He cores, in which the H envelope has been lifted off
in RLOF or driven off by wind
either before or early in the He burning the temperature and the entropy
will be slightly lower, because the insulating layer is gone, so it is not
surprising that their carbon abundance is large.
Furthermore, the core mass continually decreases because of mass loss by wind.
In fact, even for the naked
$20~\msun$ He core, corresponding to ZAMS mass $45~\msun$, 
the central carbon abundance was $\sim 33\%$ at the end of He core burning
\cite{WW95} whereas only $\sim 15\%$ is necessary for convective
carbon burning \cite{WeaW93}. For lower mass He stars the $^{12}C$
abundance was, of course, larger. 
Even with He-star wind mass loss rates reduced by half, Wellstein \&
Langer \cite{Wellstein99} find a central carbon abundance of $\gsim 1/3$
at the end of He core burning all the way up through $60~\msun$
stars, so it is clear that convective carbon
burning will take place. Unfortunately, the cores have not yet been
evolved past the CO stage.
Thus, in the range of ZAMS masses up to $\ge 60\msun$,
if the H envelope is lifted off early in the core He burning phase,
the convective carbon burning will take place after the He burning.

By ZAMS mass $\sim 40\msun$, where stars evolve into WR stars almost
independent of whether they have a companion,
the ultimate fate of the compact core is
uncertain: Brown, Weingartner \& Wijers \cite{BWW96} suggest that
1700-37, with a progenitor of about $40~\msun$ went into a low-mass
black hole. This would seem to indicate that the H-envelope of such
massive stars is blown off in an LBV phase rapidly enough that the He core
again burns as ``naked". In any case, $^{12}C$ is burned convectively
following He core burning, so the resulting Fe core should be small.

We believe that our discussion earlier in this section
indicates that single stars in the region
of ZAMS masses $\sim 20-35~\msun$ end up as high mass black holes. 
We can obtain the high mass black holes, according to our above discussion, 
if we make the He-stars burn with ``clothing", i.e., lift their H-envelope
off only following He core burning. Thus, the
evolving massive star should meet the companion main sequence star only
following He core burning (in the supergiant stage). By then its radius $R$
is several hundred $\rsun$, and its binding energy $0.6\ G M^2/R$, very small.
because of the large $R$. In order to see effects of matter stripped off
from the main sequence companion in the transient sources, we want it to
end up close to the black hole. Because of
its low binding energy the supergiant envelope will be expelled by a
relatively small binding energy of the companion, $\frac 12 M_{\rm A}
M_{{\rm B},f}/a_f$ where $a_f$ is the distance between black hole and
companion. In order to make $a_f$ small, the mass $M_\A$ of the companion
must be small.
(More massive main sequence stars will spiral in less far, hence
end up further from the black hole, and
not fill their Roche Lobes.
However, when they evolve in subgiant or giant phase they will fill it.)
Both Portegies Zwart,
Verbunt, \& Ergma \cite{PZVE97} and Ergma \& Van den Heuvel \cite{EH98}
have suggested that roughly the above region of ZAMS masses must be responsible
for the $\sim 7~\msun$ black holes in the transient X-ray sources
in order to form enough such sources.
Our scenario is essentially the same as that of de Kool et al.\cite{Kool87}
for the black hole binary A0620$-$00. We refer to this work for the
properties of the K-star companion, stressing here the evolutionary
aspects of the massive black hole progenitor.

\subsection{Calculation}

We now calculate the common envelope evolution following the formalism
of Section \ref{sec3}. Here $M_\A$ is the mass of the main
sequence companion, $M_\B$ that of the massive black hole progenitor.
The ratio
   \be
   q=\frac{M_{\A,i}}{M_{\B,i}}
   \ee
is very small and there is great uncertainty in the initial number of
binaries for such a small $q\sim 1/25$. We again take the
distribution as $dq$, and again assume $\ln a$ to be uniformly
distributed over a logarithmic interval of 7.
Again, the fraction of binaries in a given interval is
   \be
   d\phi =\frac{d (\ln a)}{7}.
   \ee

We evolve as typical
a $25 \msun$ star (B) with a companion $\sim 1 \msun$
main sequence star (star A) as the progenitor of the transient X-ray
sources. The common envelope evolution can be done as in Section \ref{sec3}.
With $M_{\B,i}=25~\msun$ and neglect of the accretion onto
the main sequence mass $M_\A$, we find from Bethe \& Brown \cite{BB98}
   \be
   \left(\frac{Y_f}{Y_i}\right)^{1.2} =
   \frac{1.2}{\alpha_{ce}}  \frac{ M_{\B,i}}{M_\A}
   \ee
where $Y=M_\B/a$. Here the coefficient of dynamical friction $c_d$ was
taken to be 6. 
The result is relatively insensitive to $c_d$, the exponent
$1.2$ resulting from $1+1/(c_d-1)$.

Thus, in our case
   \be
   \frac{Y_f}{Y_i} =
   17 \left(\frac{\alpha_{ce} M_A}{M_\odot}\right)^{-0.83} =
   30 \left(\frac{0.5}{\alpha_{ce}}\frac{M_\odot}{M_A}\right)^{0.83}.
   \ee
We expect $\alpha_{ce} \simeq 0.5$, under the assumption
that the thermal energy of the expelled envelope is equal to that
it originally possessed in the massive star (i.e.\ that it is not
used as extra energy to help remove the envelope), but it could be smaller.
{} From this we obtain
   \be
   \frac{a_i}{a_f} = \frac{M_{\B,i}Y_f}{M_{\B,f}Y_i} =
   90 \left(\frac{0.5}{\alpha_{ce}}\frac{M_\odot}{M_\A}\right)^{0.83},
   \ee
where we have taken the He star mass $M_{\B,f}$ to be $1/3$ of
$M_{\B,i}$.
In order to survive spiral-in, the final separation $a_f$ must
be sufficient so that the main sequence star lies at or inside its
Roche Lobe, about $ 0.2 a_f$ if $M_\A = \msun$.
This sets $a_f\sim 5 R_\odot = 3.5\times 10^{11}$~cm and
   \be
   a_i = 3.15
   \left(\frac{0.5}{\alpha_{ce}}\right)^{0.83}
   \times 10^{13}\; {\rm cm},
   \label{eq6.6}
   \ee
which is about 2\,AU.
This exceeds the radius of the red giant tip in the more numerous lower
mass stars in our interval, so the massive star must generally be in the {\em
supergiant} phase when it meets the main sequence star, i.e., the massive star
must be beyond He core burning.  E.g., the red giant tip (before the He core
burning) for a $20\msun$ star is at
$0.96\times 10^{13}$~cm, for a $25\msun$ star, $2.5\times 10^{13}$~cm
\cite{Schaller92}.
These numbers are, however, somewhat uncertain.
Notice that decreasing $\alpha_{ce}$ will
increase $a_i$.
Decreasing $M_A$ has little influence, because
with the smaller stellar radius the minimum $a_f$ will decrease nearly
proportionately.
Note that neglect of accretion onto the main sequence
star would change the exponent $0.83$ to unity, so accretion is
unimportant except in increasing the final mass.

Now a ZAMS $25 \msun$ star ends up at radius $6.7\times 10^{13}$~cm
($\sim 2~a_i$)
following He shell burning \cite{WZW78}.
Thus the interval between $a_i$ and $6.7\times 10^{13}$~cm
is available for spiral-in without merger
%\footnote{
%     Note that envelope removal does not occur at the original Roche Lobe 
%     on the thermal time scale $\tau_{th}$ but at the low-mass star since
%     the remaining life time of the giant
%     is $\sim 10^4$ yrs, much shorter than $\tau_{th}$.
%     }
so that a fraction
   \be
   \frac{1}{7} \ln\left( \frac{6.7}{3.15
   \left(\frac{0.5}{\alpha_{ce}}\right)^{0.83}}\right)
   \simeq 0.11
   \label{eq6.7}
   \ee
of the binaries survive spiral-in, but are close enough so that the
main sequence star is encountered by the evolving H envelope of the
massive star.
The He core burning will be completed before the supergiant has moved
out to $\sim 2$ A.U., so binaries which survive spiral-in
will have He cores which burn as ``clothed", namely as in single stars.

Given our assumptions in Section \ref{sec3},
the fraction of supernovas which arise
from ZAMS stars between 20 and $35 \msun$ is
   \be
   \frac{1}{2^{3/2}}-\frac{1}{3.5^{3/2}} =0.20
   \label{eq6.8}
   \ee
where we have assumed the mass $10 \msun$ is necessary for a star
to go supernova.
A Salpeter function with index $n=1.5$ is assumed here.
Our assumption that the binary distribution is as $dq$
is arbitrary, and gives us a factor $1/25$ for a $1~\msun$
companion.   Thus, for supernova rate 2 per century, our birth rate
for transient sources in the Galaxy is
   \be
   2\times 10^{-2}\times 0.5\times 0.11 \times 0.20 \times 0.04
   \simeq 8.8\times 10^{-6} {\rm yr}^{-1}
   \label{eq6.9}
   \ee
where $0.5$ is the assumed binarity, $0.11$ comes from eq.~(\ref{eq6.7}),
and %We comment later on
the final (most uncertain) factor $0.04$ results from a
distribution flat in $q$ and an assumed $1\msun$ companion star.

In order to estimate the number of transient sources with black holes
in the Galaxy, we should know the time that a
main sequence star of mass $ \sim 1\msun$ transfers mass to a more
massive companion. This depends on the angular-momentum loss rate
that drives the mass transfer. A guaranteed loss mechanism for
close binaries is gravitational radiation, which
for a main-sequence donor gives a mass transfer rate of
$10^{-10}\msun {\rm yr}^{-1}$, almost independent of donor mass
\cite{VerHeu95}. As mass is transferred, the mass of
the donor decreases and with it the radius of the donor.
Quite a few low-mass X-ray binaries have X-ray luminosities that imply
accretion rates in excess of $10^{-10}\msun {\rm yr}^{-1}$, leading to
suggestions of additional mechanisms for loss of angular momentum from
the binary, to increase mass transfer. Verbunt \& Zwaan \cite{VZ81} estimate
that magnetic braking can boost the transfer of mass in a low-mass binary.
We somewhat arbitrarily adopt an effective mass transfer rate of
$10^{-9} \msun {\rm yr}^{-1}$ for main sequence stars. 
In order to estimate the number of high-mass black hole, main sequence
star binaries in the Galaxy we should multiply the birth rate
eq.~(\ref{eq6.9}) by the $10^9$ yr required, at the assumed mass loss rate,
to strip the main sequence star, obtaining
8800 as our estimate.
 From the observed black-hole transient sources Wijers \cite{Wijers96} arrives
at 3000 low-mass black hole sources in the Galaxy, but regards
this number as a lower limit. With the uncertainties in formation rate
and life time, the agreement between the two numbers is as good as may be 
expected.

{
\def\arraystretch{1.0}

\begin{table}
  \caption[Black holes candidates with measured mass functions]{Parameters
   of suspected black hole binaries with measured mass functions
   \cite{Wijers96,Chen97,Bailyn98,Orosz98,Filip99,Bailyn}.
   N means nova, XN means X-ray nova. Numbers in parenthesis indicate
   errors in the last digits.
   }
   \label{tab1}
\begin{center}
\small
\newcommand{\ti}[1]{{\tiny #1}}
\noindent\begin{tabular}{@{}llccccc@{}}\hline
                &               &  compan. &$P_{orb}$   &$f(M_{X})$&$M_{opt}$ & $(l,b)$ \\
X-ray           & other         &  type    & (d)        &  ($\msun$)    &  ($\msun$)    & \\ \cline{3-7}
names           & name(s)       &  q       &$K_{opt}$   &  i          &$M_{X}$
 &    $d$ \\
                &               &  ($M_{opt}/M_X$) & (\kms)     &  (degree)   &
 ($\msun$)    &  (\kpc) \\ \hline
%------------------------------------------------------------------------------------------------------
%------------------------------------------------------------------------------------------------------
Cyg\,X-1        &\ti{V1357\,Cyg}&  O9.7Iab  &  5.5996    & 0.25(1) &  33(9)   & \ti{(73.1,$+$3.1)}\\
\ti{1956$+$350} &\ti{HDE\,226868}&           &  74.7(10)  &         &  16(5)   &
  2.5 \\ \hline
%------------------------------------------------------------------------------------------------------
LMC\,X-3        &               &  B3Ve     &  1.70      & 2.3(3)  &          & \ti{(273.6,$-$32.1)}\\
\ti{0538$-$641} &               &           & 235(11)    &         & 5.6--7.8 &
 55 \\ \hline
%------------------------------------------------------------------------------------------------------
LMC\,X-1        &               & O7--9III  &  4.22      & 0.14(5) &          & \ti{(280.2,$-$31.5)}\\
\ti{0540$-$697} &               &           & 68(8)      &         &          &
 55 \\ \hline
%------------------------------------------------------------------------------------------------------
\hline
%------------------------------------------------------------------------------------------------------
%------------------------------------------------------------------------------------------------------
XN Mon 75         &\ti{V616\,Mon}&  K4 V   & 0.3230     & 2.83-2.99      & 0.53--1.22 & \ti{(210.0,$-$6.5)} \\
\ti{A\,0620$-$003}&\ti{N Mon 1917}& 0.057--0.077 & 443(4)     & 37--44$^{(\star)}$ &  9.4--15.9 & 0.66--1.45 \\ \hline
%------------------------------------------------------------------------------------------------------
XN Oph 77       &\ti{V2107\,Oph}&  K3 V     & 0.5213     & 4.44--4.86 &  0.3--0.6 & \ti{(358.6,$+$9.1)}\\
\ti{H\,1705$-$250}&             &           & 420(30)    &   60--80   &  5.2--8.6 &  5.5: \\ \hline
%------------------------------------------------------------------------------------------------------
XN Vul 88       &\ti{QZ\,Vul}    &  K5 V   & 0.3441     &  4.89--5.13 & 0.17--0.97  & \ti{(63.4,$-$3.1)}\\
\ti{GS\,2000$+$251}&            & 0.030--0.054 & 520(16)    &  43--74   &  5.8--18.0  &  2   \\ \hline
%------------------------------------------------------------------------------------------------------
XN Cyg 89       &\ti{V404\,Cyg}&  K0 IV     & 6.4714     & 6.02--6.12  & 0.57--0.92  & \ti{(73.2,$-$2.2)}\\
\ti{GS\,2023$+$338}&\ti{N Cyg 1938, 1959}& 0.055--0.065 & 208.5(7)   &   52--60
   & 10.3--14.2   & 2.2--3.7    \\ \hline
%------------------------------------------------------------------------------------------------------
XN Mus 91       &               &  K5 V     & 0.4326     & 2.86--3.16 & 0.41--1.4 & \ti{(295.0,$-$6.1)}\\
\ti{GS\,1124$-$683}&           &  0.09--0.17 & 406(7)     &   54--65   & 4.6--8.2  & 3.0    \\ \hline
%------------------------------------------------------------------------------------------------------
XN Per 92       &               &  M0 V   & 0.2127(7)  & 1.15--1.27 & 0.10-0.97 & \ti{(197.3,$-$11.9)}\\
\ti{GRO\,J0422$+$32}&           &  0.029--0.069 & 380.6(65)  & 28--45   & 3.4--14.0 &      \\ \hline
%------------------------------------------------------------------------------------------------------
XN Sco 94       &               &  F5-G2    & 2.6127(8)  & 2.64--2.82 & 1.8--2.5
 & \ti{(345.0,$+$2.2)}\\
\ti{GRO\,J1655$-$40}&           &  0.33--0.37 & 227(2)     &   67--71   & 5.5--6.8 & 3.2   \\ \hline
%------------------------------------------------------------------------------------------------------
XN               &\ti{MX 1543-475}&  A2 V     & 1.123(8)   & 0.20--0.24 & 1.3--2.6  &  \ti{(330.9,$+$5.4)} \\
\ti{4U 1543$-$47}&                 &           & 124(4)     &  20-40    & 2.0--9.7  &  9.1(11)  \\ \hline
%------------------------------------------------------------------------------------------------------
XN Vel 93  &                & K6-M0   & 0.2852     &  3.05--3.29 & 0.50--0.65 &
                 \\
           &                & 0.137$\pm$ 0.015& 475.4(59) & $\sim$ 78 &  3.64--4.74 &        \\ \hline
%------------------------------------------------------------------------------------------------------
\end{tabular}
\end{center}
($\star$) A much higher inclination for A0620 has been claimed 
by Haswell et al. \cite{Haswell93} of up to i=70.
In this case, the lower limits on the component masses would be $M_X>3.8$ and $M_{opt}>0.22$.
\end{table}
}

\subsection{Observations}

We believe that there are many main sequence stars more massive than the
$\lsim 1~\msun$ we used in our schematic evolution, which end up further
away from the black hole and will fill their Roche Lobe only during
subgiant or giant stage. From our earlier discussion, 
we see that a $2~\msun$ main
sequence star will end up about twice as far from the black hole as
the $1~\msun$, a $3~\msun$ star, three times as far, etc. Two of the
9 systems in our Table \ref{tab1} have subgiant donors 
(V404 Cyg and XN Sco). These have the longest periods, 6.5 and 2.6 days
and XN Sco is suggested to have a relatively massive donor of $\sim 2~\msun$.
It seems clear that these donors sat inside their Roche Lobes until they
evolved off the main sequence,
and then poured matter onto the black hole once they expanded and
filled their Roche Lobe. For a $2~\msun$ star, the evolutionary time is
about a percent of the main-sequence time, so the fact that we see two subgiants
out of nine transient sources means that many more of these massive donors
are sitting quietly well within their Roche Lobes.
Indeed, we could estimate from the relative time, that there are
$2/9\times 100 =22$ times more of these latter quiet main sequence stars
in binaries.

Amazingly, this factor $22$ almost cancels the $1/25$ we had for the interval 
in $q$ over which the donors contribute. This is not coincidental.
Essentially any mass donor, at least almost up to the $25~\msun$ progenitor
of the black hole, can give rise to a common envelope phase. The BH progenitor
crosses the Herzsprung gap very quickly, in a time in which the
companion can hardly accept its mass.
(The ratio of $q\lsim 1/4$ for common envelope evolution was determined
by Kippenhahn \& Meyer-Hofmeister \cite{KipMey77} for case A mass
transfer.)
Thus, one can expect essentially all companions, up to
$q\lsim 1$, to go into common envelope evolution and contribute.
Beginning from Wijers' empirical estimate we would thus have
$(2/9)\times 100\times 3000 = 6.7 \times 10^4$ binaries with high-mass
black holes and main-sequence companions. This number is determined,
as shown above, chiefly by the number of observed systems with subgiant
donors.

If we assume that ZAMS masses $\sim 10 -18 \msun$ evolve into a neutron star,
we should have $\sim 3$ times more neutron stars than high-mass black holes
(see eq.~(\ref{eq6.8})). The range follows from our belief that SN 1987A
with progenitor $\sim 18\msun$ ZAMS went into a low-mass black hole, following
the scenario of Brown \& Bethe \cite{BB94}. On the basis of a Monte Carlo
calculation using the kick velocities of Cordes \& Chernoff \cite{CC97} we
find that $\sim 1/2$ of the binaries containing He-star, low-mass main sequence
companion (with $M\simeq 1 \msun$) will be disrupted in the explosion. Thus, we
find only a slightly higher birth rate for LMXBs (Low Mass X-ray Binaries)
with neutron stars than with
black holes, although the numbers could be equal to within our accuracy.
With comparable life times (since the donor masses and mass transfer rates
are comparable), this would give us one to a few thousand LMXBs with
neutron stars, much above the total number of observed LMXBs ($\sim130$).
Indeed, from Table 6 of Portegies Zwart \& Verbunt \cite{PZVer96}
one sees that their estimated empirical birth rate for low-mass X-ray binaries
is $2\times 10^{-7}$ yr$^{-1}$, whereas in either theoretical evolution
including kick velocities they obtain $4\times 10^{-6}$ yr$^{-1}$. This
factor of 20 discrepancy is by far the greatest between theoretical
and empirical rates in their table, and supports our point that many
of the neutron stars must have disappeared along the way. Alternatively,
a large number of LMXBs with neutron stars could be transients as well
(like, e.g.\ Aql\,X-1).
Just at the present there are new developments in the evolution of
low-mass X-ray binaries, which we shall shortly summarize in Section~\ref{sec7}.

As we showed below eq.~(\ref{eq6.6}),
the He core of the massive star will in general
be uncovered only after He core burning is completed.  The remaining time for
He burning (in a shell) will be short, e.g., for a $20 \msun$ ZAMS star it is
only $1.4 \times 10^4$ years \cite{Schaller92}.  Therefore the mass loss by
wind after uncovering the He core will not be large, and when the star finally
becomes a supernova, its mass will be almost equal to the He core of the
original star.  The latter can be calculated from
   \be
   M_{\rm He} \simeq 0.10 \ (M_{\rm ZAMS})^{1.4}
   \label{eq6.10}
   \ee
so for ZAMS masses $20-35\msun$ $M_{\rm He}$ will lie in the interval
$\sim 7- 14~\msun$.

Bailyn et al. \cite{Bailyn98} find the black hole masses in transient sources
to be clustered about $\sim 7\msun$, except for V404 Cyg which has a
higher mass. This is in general agreement with our scenario, because
most of the black holes will come from the more numerous stars of ZAMS
mass not far from our lower limit of $\sim 20\msun$.
Two points are important to note:

\begin{enumerate}

\item Not much mass can have been lost by wind. Naked He stars have
rapid wind loss.
However in our scenario the He star is made naked only
during He shell burning and therefore does not have much time
($\lsim 10^4$ yr) to lose mass by wind.

\item There are good reasons to believe that the initial He core will be
rotating \cite{Minesh93}. The way in which
the initial angular momentum affects the accretion process has been studied
by Mineshige et al. \cite{Minesh97} for black hole accretion in supernovae.
In general accretion discs which are optically thick and advection
dominated are formed. The disc is hot and the produced energy and photons
are advected inward rather than being radiated away. The disc material
accretes into the black hole at a rate of $> 10^6 {\dot M}_{\rm Edd}$ for
the first several tens of days. Angular momentum is advected outwards.
Our results show that little mass is lost, because the final
$\sim 7~\msun$ black hole masses are not much less massive than the He core
masses of the progenitors, and some mass is lost by wind before the
core collapses.
The latter  loss will not, however, be great, because there is not much
time from the removal of the He envelope until the collapse.

\end{enumerate}

Accretion of the He into the black hole will differ quantitatively
from the above, but we believe it will be qualitatively similar.
The fact that the helium must be advected inwards and that little mass is lost
as the angular momentum is advected outwards is extremely
important to establish. This is because angular momentum, essentially
centrifugal force, has been suggested by Chevalier \cite{Chevalier96} to hold
up hypercritical accretion onto neutron stars in common envelope evolution.
(Chevalier \cite{Chevalier93} had 
first proposed the hypercritical accretion during
this evolutionary phase to turn the neutron stars into black holes,
the work followed up by Brown \cite{Brown95} and Bethe \& Brown \cite{BB98}.)
However, once matter is advected onto a neutron star, temperatures
$\gsim 1$ MeV are reached so that neutrinos can carry off the energy.
The accreted matter simply adds to the neutron star mass, evolving
into an equilibrium configuration. Thus, this accretion does not differ
essentially from that into a black hole.
In either case of neutron star or black hole an accretion disc or
accretion shock, depending on amount of angular momentum, but both
of radius $\sim 10^{11}$~cm, is first formed, giving essentially
the same boundary condition for the hypercritical accretion 
for either black hole or neutron star. Thus, the masses of the
black holes in transient sources argue against substantial inhibition
of hypercritical accretion by jets, one of the Chevalier suggestions
\cite{Chevalier96}.

Measured mass functions, which give a lower limit on the black hole
mass are given in Table~\ref{tab1}. Only GRO J0422+32 and 4U 1543-47
have a measured mass function $\lsim 3  \msun$.  Results of Callanan et
al. \cite{Callanan96} indicate that the angle $i$ between the orbital
plane and the plane of the sky for GRO J0422+32 is $i<45^\circ$, and
recent analysis \cite{Orosz98} indicate that the angle $i$ for 4U 1543-47
is $20^\circ <i< 40^\circ$.  So both GRO J0422+32 and 4U 1543-47 also
contain  high-mass black holes.

Based on the observations of Kaper et al. \cite{Kaper95} that the
companion is a hypergiant, Ergma \& Van den Heuvel \cite{EH98} argue that
the progenitor of the neutron star in 4U1223-62 must have a ZAMS mass
$\gsim 50\msun$.  Brown, Weingartner \& Wijers \cite{BWW96}, by similar
argumentation, arrived at $\sim 45\msun$, but then had the difficulty
that 4U1700-37, which they suggested contains a low-mass black hole,
appeared to evolve from a lower mass star than the neutron star in 1223.
Wellstein \& Langer \cite{Wellstein99} suggest the alternative that in
1223 the mass occurs in the main-sequence phase (Case A mass transfer),
which would be expected to be quasi conservative.  They find that the
progenitor of the neutron star in 1223 could then come from a mass as low
as $26\msun$. This is in agreement with Brown et al. \cite{BWW96} for
conservative mass transfer (their Table 1), but these authors discarded
this possibility, considering only Case B mass transfer in which
case considerable mass would be lost.

Wellstein \& Langer \cite{Wellstein99} 
are in agreement with Brown et al. \cite{BWW96} that
4U1700-37 should come from a quite massive progenitor.
Conservative evolution here is not possible because of the short period
of 3.4 days \cite{Rubin96}. The compact object mass is here
$1.8 \pm 0.4\msun$ \cite{Heap92}. Brown et al. \cite{BWW96} suggest
that the compact object is a low-mass black hole. The upper mass limit
for these was found by Brown \& Bethe \cite{BB94} to be $\sim 1.8\msun$,
as compared with an upper limit for neutron star masses of $\sim 1.5\msun$.
Thus, there seems to be evidence for some ZAMS masses of $\sim 40-50\msun$
ending up as low-mass compact objects, whereas we found that lower mass
stars in the interval from $\sim 20-35\msun$ ended up as
high-mass black holes. In this sense we agree with 
Ergma \& Van den Heuvel~\cite{EH98} 
that low-mass compact object formation ``is connected with
other stellar parameters than the initial stellar mass alone". We suggest,
however, following Brown et al. \cite{BWW96} that stars in binaries evolve
differently from single stars because of the different evolution of the
He core in binaries resulting from RLOF in their evolution. Namely,
``naked" He cores evolve to smaller final compact objects than ``clothed"
ones.

In fact, this different evolution of binaries was found by Timmes, Woosley
\& Weaver \cite{TWW96}. They pointed out that stars denuded of their hydrogen
envelope in early RLOF in binaries would explode as Type Ib supernovae.
They found the resulting remnant gravitational mass following explosion
to be in the interval of $1.2-1.4\msun$, whereas in exploding stars of all
masses with hydrogen envelope (Type II supernova explosion) they found
a peak at about $1.28\msun$, chiefly from stars of low masses and another
peak at $1.73\msun$ more from massive stars. Our Fe core masses in
Fig.~\ref{fig1} come from essentially the same calculations, but the
``Remnant" masses
of Woosley \& Weaver \cite{WW95} are somewhat greater than those used
by Timmes et al. \cite{TWW96}. In fact, the differences between the masses we
plot and those of Timmes et al. come in the region $\sim 1.7-1.8\msun$
(gravitational). This is just in the Brown \& Bethe \cite{BB94} range for
low-mass black holes. It may be that some of the stars with low-mass
companions evolve into low-mass black holes. Presumably
these would give lower luminosities than the high-mass black holes, although
at upper end of the mass range we discuss 4U1700-37 seems to be an
example of such a system.
Of course here the high luminosity results from the high mass loss rate
of the giant companion.
There are substantial ambiguities in fallback, etc., from the explosion.
Our point in this paper is that most of the higher mass single stars
$20-35\msun$ go into high mass black holes. (The Brown \& Bethe \cite{BB94}
limit for low-mass black hole formation is $\sim 1.5-1.8\msun$ gravitational,
but there is some give and take in both lower and upper limit.
Also the stars are not all the same. In particular different metallicities
will give different wind losses.)

%----------------------------------------------------------
\section{Evolution of Low Mass X-ray Binaries}
\label{sec7}

We shall briefly point out new developments in the evolution of
low-mass X-ray binaries. These were foreseen in the excellent review
by Van den Heuvel \cite{Heuvel95}, and there has been substantial
development in this field lately.

Low mass X-ray binaries are considered to be progenitors of recycled 
pulsars with helium white dwarf companions.
In order to bring the magnetic fields of the latter down to
$\sim 10^8$ G and to speed them up to their final period, Van den Heuvel
\& Bitzaraki \cite{HeuBit95} had the neutron star accreting $\sim 0.5~\msun$
from the main-sequence donor. More detailed recent calculations by Tauris
\& Savonije \cite{TauSav99} find that if the initial orbital period is
below $\sim 30$ days with a main sequence donor of $\sim 1~\msun$
which undergoes stable mass transfer with the neutron star, the mass of
the latter is increased up to $\sim 2~\msun$ if the amount of material
ejected as a result of propeller effect or disk instabilities
is insignificant. This presents a problem for
us because the Brown \& Bethe \cite{BB94} mass limit for neutron stars is
$1.5~\msun$. From this limit, we would say that these neutron stars in 
low-mass X-ray binaries would have gone into black holes.

A way out of this problem was suggested by Van den Heuvel \cite{Heuvel95},
which is called the evolution of Her X-1 type X-ray binaries
(see especially the Appendix of Van den Heuvel \cite{Heuvel95}).
In this case a radiative donor more massive than the neutron star pours
matter in unstable mass transfer across the Roche Lobe onto the neutron 
star. This mass transfer can occur onto the accretion disc by as much as
$\sim 10^4\ \dot M_{\rm Edd}$, 
if $\dot M_{\rm Edd}\sim 1.5\times 10^{-8}~\msun$ yr$^{-1}$
is accreted onto the neutron star, since the Eddington limit goes linearly
with $R$ and the radius of the disc can be $\sim 10^{10}$ cm.
The advection dominated inflow-outflow solution (ADIOS) of Blandford
\& Begelman \cite{BlandBegel99} suggests that the binding energy of the matter
released at the neutron star can carry away mass, angular momentum and
energy from the gas accreting onto the edge of the accretion disc provided
the latter does not cool too much. In this way the binding energy of
a gram of gas at the neutron star can carry off $\sim 10^3$ grams of gas
at the edge of the accretion disc. Such radiatively-driven outflows are
suggested by King \& Begelman \cite{KingBegel99} to enable common
envelope evolution to be avoided. Tauris \& Savonije \cite{TauSav99}
have carried out a detailed evolution of low-mass X-ray binaries with
$P_{\rm orb}>2$ days using computer programs based on Eggleton's,
which for radiative and semiconvective donors follow, in at least
a general way, the above ideas.
For a deeply convective donor a short phase of rapid mass loss may
reach a rate as large as
$10^4 \dot M_{\rm Edd}$ while the mass of the donor drops to well
below the neutron star mass. Although rates $>10^4~{\dot M}_{\rm Edd}$ 
would be hypercritical for spherical accretion, somewhat higher rates
survive hypercritical accretion provided angular momentum is taken 
into account \cite{Chevalier96}. The important point is that the donor
mass can be brought down sufficiently far before stable mass transfer
at a rate $\lsim \dot M_{\rm Edd}$ sets in, so that the neutron star
can avoid accreting sufficient mass to send it into a black hole.
It is not clear what percentage of the neutron stars will survive black-hole
fate. Our rough estimates in Section \ref{sec6} indicate that only
a small fraction need to do so.

For even more massive donors (2--6\,$\msun$)
which are either radiative or semiconvective,
work by Tauris et~al.\ \cite{Tauris99} indicates that the low-mass
X-ray binaries with C/O white dwarf (CO) companions can be made
in much the same way. In an earlier paper, Van den Heuvel \cite{Heuvel94}
had suggested
that these binaries would originate from donor stars on the asymptotic
giant branch. In order to evolve these, he needed an efficiency
$\alpha >1$; i.e., sources additional to those included in our earlier
common envelope evolution, such as mass loss by instabilities in the AGB,
dissociation energy, etc. have to participate in helping to remove the
envelope of the donor star.

King and Ritter \cite{KingRitter99} have computed a scenario for Cyg\,X-2 
with an initial donor mass of $\sim3.6\msun$. Currently the donor has
a mass of $0.5-0.7\msun$ and a large radius, about $7\rsun$. About $2\msun$
must have been lost in super-Eddington accretion, roughly along the lines
sketched above. More massive donors can lead to relatively more massive
white dwarf companions, which will be C/O white dwarfs.

In fact, the present situation is that no circular NS-CO$_c$\footnote{
     The lower suffix $c$ ($e$) denotes the circular (eccentric)
     binaries.
     }
binaries
which went through common envelope evolution seem to be observed,
the alternative Tauris et al.\cite{Tauris99} evolution which avoids
common envelope evolution being preferred.
This presents a real dilema for the standard scenario of common
envelope evolution. It seems clear \cite{KerkKul99} that in the
binary B2303$+$46 the companion to the pulsar is a C/O white dwarf.
B2303$+$46 is an eccentric binary NS-CO$_e$, indicating that the
neutron star was formed last. This is confirmed by the unrecycled
field strength of the pulsar $B=8\times 10^{11}$ G. Cases have made that 
the recently discovered J1141$-$6545 \cite{TauSen99} and B1820$-$11
\cite{PZYun99} are also NS-CO$_e$ binaries.

On the other hand, evolutionary calculations show that formation
probability of NS-CO$_c$ binaries through common envelope evolution
is $\gsim 50\%$ as probable as of NS-CO$_e$ binaries \cite{BLPZ99}.
In this evolution the pulsar magnetic moment will be recycled,
brought down at least a factor of 100 \cite{Brown95} and possibly
even further, down to the empirical values of $\sim 5\times 10^8$ G
found in the NS-CO$_c$ binaries. The lowering of the magnetic fields
increases the time of observation by a factor of $\sim 100$ or
of $2000$, depending on whether the theoretical or empirical
magnetic field is used. Since we fairly certainly observe at least
one NS-CO$_e$ binary, we should see either 100 or 2000 NS-CO$_c$
binaries which have gone through common envelope evolution. We
certainly don't see anything like this, at most the 5 that had
earlier been attributed to common envelope evolution, and probably none.
Brown, Lee, \& Portegies Zwart \cite{BLPZ99} remove at least most
of this discrepancy by showing that with the introduction of
hypercritical accretion the neutron star in common envelope evolution
with the evolving main sequence companion goes into a black hole.

%----------------------------------------------------------
\section{Discussion and Conclusion} 
\label{sec8}

Our chief new point in the evolution of binaries of compact objects
is the use of hypercritical accretion in common envelope evolution,
although the idea of hypercritical accretion is not new (Section \ref{sec3}).
Chevalier \cite{Chevalier96} discussed the possibility that angular
momentum might hinder hypercritical accretion.
In his treatment of the accretion disc, he assumed gas pressure to
dominate, in order to raise the temperature sufficiently for neutrinos
to be emitted. This entailed a tiny viscosity, characterized by
$\alpha \lsim 10^{-6}$ in the $\alpha$-description. More reasonable
values of $\alpha$ are $\sim 0.1$.

Bethe, Brown, \& Lee \cite{BBL99} have shown that for larger $\alpha$'s,
$\alpha \sim 0.01 - 1$, the disc pressure is radiation dominated, and they
find a simple hypercritical advection dominated accretion flow (HADAF)
of matter onto the neutron star.

The Bethe, Brown, \& Lee HADAF appears to reproduce the Armitage
\& Livio \cite{Armit99} numerical two-dimensional hydro solution.
These latter authors suggest that jets will prevent hypercritical
accretion by blowing off the accreting matter. At such high rates of
accretion $\sim 1\msun$ yr$^{-1}$ the Alfven radius is, however,
close to the neutron star surface, and we believe that this will
effectively, shut down any magnetically driven jets.

In Section \ref{sec7} we discussed the advection of a rotating He
envelope into a black hole. We believe that two possibilities exist.
Phinney \& Spruit suggest \cite{PhiSpr98} that the magnetic turbulence
is strong enough to keep the He envelope in corotation with the core
of the star until shortly before it evolves into a black hole.
Then not much angular momentum would have to be advected away in order
to let the matter accrete. Alternatively, magnetic turbulence is strong
enough so that angular momentum can be carried away from a rapidly
rotating He core; then the matter can accrete. From the measured masses
of $\sim 7\ \msun$ we know that most of the He core must fall into the
black hole, so one of these scenarios should hold. Both favor high
magnetic turbulence, lending credence to the Chevalier suggestion
we quoted.

%-----------------------------------------------------------------------
\section*{Appendix}

\addcontentsline{toc}{section}{Appendix}

\appendix
\section{Common Envelope Evolution of Cygnus X-3}

The closeness of the compact object in Cyg\,X-3 to its $\sim 10\ \msun$
companion helium star bears witness to an earlier stage of common
envelope evolution. Although the mass of the He star has not been measured,
the star is similar to V444 Cygni, the mass of which is 
$9.3\pm 0.5\ \msun$~\cite{Marchenko94}. For example, from the period change
its mass loss rate would be 
$\dot M_{\rm dyn} =0.6\times 10^{-5} 
(M_{\rm He}/10\ \msun)\ \msun\  {\rm yr}^{-1}$ 
\cite{Kerk96} whereas that of V444 Cygni is $\dot M_{\rm dyn}=1\times 10^{-5}
\ \msun$ yr$^{-1}$ \cite{Khaliullin84} indicating an $M_{\rm He}
\sim 10~\msun$.
Mass loss rates cannot easily be obtained from W.-R. winds because
of large nonlinear effects which necessitate corrections for
``clumpiness". However polarization measurement of the Thomson scattering,
which depend linearly on the wind, give a mass loss rate of $\sim \dot M
=0.75\times 10^{-5}\ \msun$ yr$^{-1}$ \cite{Louis93}, roughly compatible
with the period change. In agreement with many other authors we take 
$M_{\rm He}=10\ \msun$ in Cyg\,X-3. 

Here we evolve a massive O-star binary with initial ZAMS masses of 
$33\ \msun$ and $23\ \msun$ as possible progenitor for Cyg\,X-3. 
In red giant phase the $33\ \msun$ star will transfer its H envelope to
the $23\ \msun$ companion, leaving a He star of 
   \be
   M_{\rm He}=0.1\; M_{\rm ZAMS}^{1.4} =13\;\msun .
   \ee
With efficiency of mass transfer assumed to go as $q^2$, about half of 
the $20\;\msun$ H-envelope will be accepted by the companion, which then
becomes a rejuvenated $33\;\msun$ star. The He core of the primary then
explodes, going into a $1.5\;\msun$ compact object, neutron star or low-mass
black hole. After the companion $33\;\msun$ star evolves, the binary
will go into common envelope evolution. Eq.~(\ref{eq3.22}) can be written
   \be
   \left(\frac{Y_f}{Y_i}\right)
   =\left(\frac{2.4 \; M_{\B,i}}{M_{\A,i}}\right)^{\frac{c_d-1}{c_d}}
   \ee
where we again take $c_d=6$. With $M_{\B,i}=33\;\msun$ and 
$M_{\A,i}=1.5\;\msun$,
   \be
   Y_f/Y_i =27.
   \ee
The compact object mass scales as
   \be
   M_\A \propto Y^{1/(c_d-1)} =Y^{1/5}
   \ee
so that
   \be
   M_{\A,f}=2.9 \;\msun
   \ee
and the final compact object is certainly a black hole, in agreement with
Cherepaschchuk \& Moffat~\cite{Cherep94} and Ergma \& Yungelson\cite{ErgYu98}.
We believe our evolution here to show that this $\sim 3\;\msun$ black hole is
about the most massive that can be formed in common envelope evolution
by accretion onto a low-mass compact object, since our $33\;\msun$ companion is
near to the ZAMS mass range that will lose mass in an LBV phase, unsuitable
for common envelope evolution, so
it cannot be made much more massive. We next find
   \be
   \frac{a_i}{a_f}=\frac{M_{\B,i}}{M_{\B,f}}\;\frac{Y_f}{Y_i} \simeq 70.
   \ee
For an $a_f\sim 3.5\;\rsun$ this gives
   \be
   a_i \sim 250 \;\rsun
   \ee
comfortably within the red giant range.

Following Ergma \& Yungelson \cite{ErgYu98} we calculate the accretion rate as
   \be
   \dot M_{\rm acc}=0.14 \left(\frac{M_{\rm BH}}{\msun}\right)^2
   v_{1000}^{-4}\; P_{{\rm hr}}^{-4/3}
   \left(\frac{\msun}{M_{\rm tot}}\right)^{2/3} \dot M_{\rm wind}.
   \ee
Here $v_{1000}$ is the wind velocity in units of 1000 km s$^{-1}$ and 
$P_{\rm hr}$ is the period in hours.\footnote{ 
    Through a slip, the two factors preceding $\dot M_{\rm wind}$
    appear in the denominator in \cite{ErgYu98}, although we confirm that they
    carried out their calculations with the correct formula.}
For $\dot M_{\rm wind}$ we, as Ergma \& Yungelson, take $\dot M_{\rm dyn}$.
These authors take $v_{1000}=1.5$, essentially the result of 
Van Kerkwijk et al. \cite{Kerk96}. An earlier estimate by Van Kerkwijk
et al.~\cite{Kerk93} was $v_{1000}=1$. We believe that the $v_{\rm wind}$ to
be used here may be different from the (uncertain) measured terminal
wind velocities, because the velocity near the compact object is substantially
less.  Therefore, we take $v_{1000}=1.$
Taking $\dot M_{\rm wind}=\dot M_{\rm dyn}$ we obtain
   \be
   \dot M_{\rm acc}=2.2\times 10^{-7} \;\msun\; {\rm yr}^{-1}
   \ee
This is to be compared with
   \be
   \dot M_{\rm Edd}=4\pi c R/\kappa_{\rm es}
   =2.6\times 10^{-8} \; (M_{\rm BH}/\msun) \;\msun \; {\rm yr}^{-1}
   \ee
where $\kappa_{\rm es}=0.2 \;{\rm g/cm}^2$ for He accretion. 
Our result is in fair
agreement with Ergma \& Yungelson \cite{ErgYu98}, who find
$\dot M_{\rm Edd}\sim 2.3\times 10^{-7}\;\msun$ yr$^{-1}$ for a $10\;\msun$ 
black hole. The presence of jets in Cyg\,X-3 argues for super Eddington rates
of accretion, which we find.

Cherepaschchuk and Moffat \cite{Cherep94} estimated the total luminosity
of Cyg\,X-3 to be $L_{\rm bol}\sim 3\times 10^{39}$ erg. The efficiency
of black hole accretion varies as
   \be
   0.057 < \epsilon < 0.42
   \ee
for a black hole at rest to a (maximally rotating) Kerr black hole. We
expect the black hole to be spun up by accretion from the wind or accretion
disc. Taking an intermediate $\epsilon =0.2$, we find
   \be
   L=2.5\times 10^{39}\; {\rm erg\; s}^{-1}
   \ee
in rough agreement with the Cherepaschchuk and Moffat value.

Cyg\,X-3 is often discussed as the ``missing link" in binary pulsar
formation. In fact, because of its high He star mass, upon explosion of
the latter,  it most probably will break up. But it should be viewed as
``tip of the iceberg" \cite{Heuvel95}, in that there must be a great
many more such objects with lower mass He stars which are not seen. We
have shown, in section \ref{sec3} however, that these objects
are more likely to contain a black hole than a neutron star.

In our evolutionary scenario, the He star progenitor has about the same
ZAMS mass as that of the primary. Thus, the fate of the ``naked" He
star should be the same low-mass compact object, neutron star or low-mass
black hole that resulted from the explosion of the primary.

%-----------------------------------------------------------
%\setcounter{equation}{0}
%\renewcommand{\theequation}{B.\arabic{equation}}
%\renewcommand{\thesubsection}{B}
\section{Implications for LIGO}

Our results that there are 10 times more 
\footnote{ Actually about 20 times more if we include the binaries in which
           the pulsar goes into a black hole in the He shell burning evolution.
           However, these will have masses not very different from the binary
           neutron stars so we do not differentiate them. }
black hole, neutron star binaries than binary neutron stars has important
results for LIGO, the detection rates of which were based on the
$\sim 10^{-5}$ per year per galaxy rates of merging for the latter. The
combination of masses which will be well determined by LIGO is the chirp
mass
   \be
   M_{\rm chirp}=\mu^{3/5} M^{2/5}
    = (M_1 M_2)^{3/5} \ (M_1+M_2)^{-1/5}
   \ee
where $M=M_1+M_2$ is the total system mass. The chirp mass of a NS-NS
binary, with both neutron stars of mass $1.4 \;\msun$, is $1.2 \;\msun$.
A $10^{-5}$ birth rate implies a rate of 3 yr$^{-1}$ out to
200 Mpc \cite{Phinney91}. Kip Thorne informs us that LIGO's first
long gravitational-wave search in 2002-2003 is expected to see binaries
with $M_{\rm chirp}=1.2 \;\msun$ out to 21 Mpc. 

The chirp mass corresponding to the Bethe \& Brown \cite{BB98}
LMBH-NS binary with masses $2.4 \;\msun$ and $1.4\;\msun$ is $1.6\;\msun$.
Including an $\sim 30$ \% increase in the rate to allow for high-mass
black-hole, neutron-star mergers (which should be regarded as a lower
limit because of the high-mass limit of $80\;\msun$ used by Bethe \& Brown
for going into a HMBH) gives a 26 times higher rate than Phinney's
estimate for NS-NS mergers. There factors are calculated from the
signal to noise ratio, which goes as $M_{\rm chirp}^{5/6}$ and then
cubing it to obtain the volume of detectability. We then predict
a ratio of $3\times (21/200)^3\times 26 =0.09$ yr$^{-1}$ for 2003,  rather
slim. The enhanced LIGO interferometer planned to begin in 2004 should reach
out beyond 150 Mpc for $M_{\rm chirp}=1.2\;\msun$, increasing the detection
rate to $3\times (150/200)^3\times 26=33$ yr$^{-1}$. We therefore predict
that LIGO will see more mergers per month than NS-NS mergers per year.

%-----------------------------------------------------------
%\setcounter{equation}{0}
%\renewcommand{\theequation}{C.\arabic{equation}}
%\renewcommand{\thesubsection}{C}
\section{Binary Contributions to Gamma Ray Bursters}

The sheer numbers of black-hole, neutron-star  binaries should dominate
the mergers for gravitational waves, which could be detected by LIGO
\cite{BWLLB99}.
For gamma-ray bursts, the presence of an event horizon eases the baryon 
pollution problem, because energy can be stored in the rotational energy
of the black hole, and then released into a cleaner environment via the
Blandford-Znajek magnetohydrodynamic process.

Binaries containing a black hole, or single black holes, have been
suggested for some time as good progenitors for gamma-ray bursts
\cite{Paczynski91,Paczynski98,Mochko93,Woosley93,FryWoos98,MacWoos98}.
Reasons for this include
the fact that the rest mass of a stellar mass black hole is comparable
to what is required to energize the strongest GRB. Also, the horizon
of a black hole provides a way of quickly removing most of the material
present in the cataclysmic event that formed it. This may be important
because of the baryon pollution problem: we need the ejecta that give
rise to the GRB to be accelerated to a Lorentz factor of 100 or more,
whereas the natural energy scale for any particle near a black hole is
less than its mass. Consequently, we have a distillation problem of taking
all the energy released and putting it into a small fraction of the total
mass. The use of a Poynting flux from a black hole in a magnetic 
field \cite{BZ77} does not require the presence of much mass,
and uses the rotation energy of the black hole, so it provides naturally
clean power.

As a neutron star in a binary moves nearer to a black hole companion, it
is distorted into a torus around the latter. Most of the torus matter
enters the black hole from the last stable Keplerian orbit of
$R=6 GM_{\rm BH}/c^2$, carrying considerable angular momentum. In the
process the black hole is spun up until it rotates with some fraction of the
speed of light. A magnetic field which originates from the neutron star,
but which could have been enhanced by differential rotation is anchored
in the remaining part of the torus, the accretion disc.

When a rapidly rotating black hole is immersed in a magnetic field, frame
dragging twists the field lines near the hole, which
causes a Poynting flux to be emitted from near the black hole. This is
the Blandford-Znajek mechanism \cite{BZ77}. The source of energy for the flux
is the rotation of the black hole. The source of the field is the surrounding
accretion disk or debris torus. We showed \cite{LWB99}
that at most 9\% of the rest mass of a rotating black hole can be
converted to a Poynting flux, making the available energy for powering
a GRB
\be
   E_{{\rm BZ}}=1.6\times 10^{53}\; (M/M_\odot)\; {\rm erg}.
\label{eq:ebz}
\ee
The power depends on the applied magnetic field:
\be
   P_{{\rm BZ}}\sim 6.7\times 10^{50}\ B_{15}^2 \
      (M/M_\odot)^2\ {\rm erg\; s}^{-1}
\label{eq:pbz}
\ee
(where $B_{15}=B/10^{15}$\,G). This shows that modest variations in the
applied magnetic field may explain a wide range of GRB powers, and therefore
of GRB durations. There has been some recent dispute in the literature
whether this mechanism can indeed be efficient \cite{Li99} and whether the
power of the BH is ever significant relative to that from the disk
\cite{LiOgP99}. The answer in both cases is yes,
as discussed by Lee, Wijers, \& Brown \cite{LWB99}.

The issue, therefore, in finding efficient GRB sources among black holes
is to find those that spin rapidly. There are a variety of reasons why
a black hole might have high angular momentum. It may have formed from
a rapidly rotating star, so the angular momentum was there all along
(`original spin', according to Blandford \cite{Bland99}); 
it may also have accreted
angular momentum by interaction with a disk (`venial spin') or have
formed by coalescence of a compact binary (`mortal spin').  We shall
review some of the specific situations that have been proposed in turn.

Neutron star mergers are among the oldest proposed cosmological GRB sources
\cite{ELPS89,gdn:87,Paczy86},
and especially the neutrino flux is still actively studied as a GRB power
source \cite{ruffert}.
However, once the central mass has collapsed to
a black hole it becomes a good source for BZ power, since it naturally
spins rapidly due to inheritance of angular momentum from the binary
\cite{RM92}. Likewise BH-NS binaries \cite{LS74} will rapidly transfer a
large amount of mass once the NS fills its Roche lobe, giving a rapidly
rotating BH \cite{KL98}. The NS remnant may then
be tidally destroyed, leading to a compact torus around the BH. It
is unlikely that this would be long-lived enough to produce the longer
GRB, but perhaps the short ($t\lsim1$\,s) ones could be produced 
\cite{FWH99}. However, mass transfer could stabilize
and lead to a widening binary in which the NS lives until its mass drops
to the minimum mass of about $0.1\;\msun$, and then becomes a debris torus
\cite{simon2} By then, it is far enough away that the resulting
disk life time exceeds 1000\,s, allowing even the longer GRB to be made.
Thus BH-NS and NS-NS binaries are quite promising. They have the added
advantage that their environment is naturally reasonably clean, since
there is no stellar envelope, and much of the initially present baryonic
material vanishes into the horizon.

In addition to the mergers from compact objects, Fryer \& Woosley
\cite{FryWoos98} suggested that GRBs could originate from the coalescence of
low-mass black hole and helium-star binaries in the Bethe \& Brown
\cite{BB98} scenario. From eq.~(\ref{eq3.28}) we see that binaries 
survived in the initial range of $0.5\times 10^{13} {\rm cm}<a_i<
1.9\times 10^{13}{\rm cm}$. Inside that range for $0.04\times 10^{13}{\rm cm}
< a_i < 0.5\times 10^{13}$ cm the low-mass black hole coalesces with the
core. Hence, using a separation distribution flat in $\ln a$, coalescences 
are more common than low-mass black-hole, neutron-star binaries by a factor
$\ln (0.5/0.04)/\ln (1.9/0.5) =1.9$. In Bethe \& Brown \cite{BB98}
the He star compact-object binary was disrupted $\sim 50$ \% of the time
in the last explosion, which we do not have here. Thus, the rate of low-mass
black-hole, He-star mergers is 3.8 times the formation rate of low-mass
black-hole, neutron-star binaries which merge, or $R=3.8\times 10^{-4}$ 
yr$^{-1}$ in the Galaxy.

In Table~\ref{ta:rates} we summarize the formation rates of GRBs and
gravity waves from the binaries considered in this review.

\renewcommand{\thetable}{2}
\begin{table}
\caption{Summary of the formation rates of various sources of 
         gamma-ray bursts (GRB) or gravity waves (GW) from
         the binaries considered in this review.
         L(H)BH means low- (high-)mass black hole.}
\label{ta:rates}
\begin{tabular}{lccr} \hline
object                       & GRB & GW & rate [GEM$^\star$] \\ \hline
NS - NS merger               &  X  & X  &  10 \\
NS - BH merger               &  X  & X  & 100 \\
WR star - LBH merger         &  X  &    & 380 \\ \hline
\end{tabular}\\
{$^\star$
GEM means Galactic Events per Megayear; rates are quoted for redshift 0.}
\end{table}

Because gamma-ray bursts have a median redshift of 1.5--2 \cite{wbbn:98}), 
and the supernova rate at that redshift was 10--20 times
higher than now, the gamma-ray burst rate as observed is higher than
one expects using the above rates. However, for ease of comparison
with evolutionary scenarios we shall use the GRB rate at the present
time (redshift 0) of about 0.1\,GEM. (Wijers et~al. \cite{wbbn:98} found a
factor 3 lower rate, but had slightly underestimated it because they
overestimated the mean GRB redshift; see ref.~\cite{FWH99}
for more extensive discussions of the redshift dependence).
An important uncertainty is the
beaming of gamma-ray bursts: the gamma rays may only be emitted in narrow
cones around the spin axis of the black hole, and therefore most GRBs may
not be seen by us. An upper limit to the ratio of undetected to detected
GRB is 600 \cite{mrw:99}, so an upper limit to the
total required formation rate would be 60\,GEM. We may have seen beaming
of about that factor or a bit less in GRB\,990123 \cite{kulka:99},
but other bursts (e.g.\ 970228, 970508) show no evidence of beaming in
the afterglows (which may not exclude beaming of their gamma rays).
At present, therefore, any progenitor with a formation rate of
10\,GEM or more should be considered consistent with the observed GRB
rate.

An exciting possibility for the future will be to receive both 
gravitational-wave and gamma-ray burst signals from the same merger, with
attendant detailed measurement, which would give witness to them arising
from the same binary.

Because we dealt in this review with binaries, we did not explain one
popular model of GRBs, the Woosley Collapsar model \cite{Woosley93}.
In this model a black hole is formed in the center of a rotating 
W.-R. star. The outer matter can then be accreted into the neutron star,
spinning it up. 
If, however, magnetic turbulence is sufficient to keep the envelope of the
progenitor in corotation with the core until a few days before collapse
of the latter, as suggested by Phinney \& Spruit \cite{PhiSpr98}
the He envelope could not furnish enough angular momentum to the black
hole for the latter to drive the necessary jets
(see end of Section \ref{sec8}).

%-----------------------------------------------------------------------
\section*{Acknowledgements}

G.E.B. and C.H.L. wish to acknowledge support from  
the U.S. Department of Energy under Grant No.  DE--FG02--88ER40388.
We thank C. Bailyn for his help with Table~1 and discussions on 
black-hole transients.

%-----------------------------------------------------------------------

\end{document}